# Synthesis and structure of $SrCo_2Si_2$ and $CaRh_2Si_2$ - isoelectronic variants of the parent superconductors $\mathit{Ae}Fe_2As_2$ and study of the influence of the valence electron count in $CaFe_{2-x}Rh_xSi_2$

**Viktor Hlukhyy*, Andrea V. Hoffmann, Thomas F. Fässler**

Department Chemie, Technische Universität München, Lichtenbergstr. 4, D-85747 Garching, Germany

* Corresponding author:
Dr. Viktor Hlukhyy, Department Chemie, Technische Universität München, Lichtenbergstr. 4, D-85747 Garching, Germany
E-mail: viktor.hlukhyy@lrz.tu-muenchen.de

## Abstract

The finding of superconductivity in $Ba_{0.6}K_{0.4}Fe_2As_2$ put the attention on the investigation of compounds that crystallize with $ThCr_2Si_2$ structure type such as $AT_2X_2$ ($A$ = alkali/ alkaline earth/ rare earth element; $T$ = transition metal and $X$ = element of the 13-15$^{th}$ group) . In this context the silicides $CaFe_2Si_2$, $CaFe_{0.68(6)}Rh_{1.32(6)}Si_2$, $CaRh_2Si_2$ and $SrCo_2Si_2$ have been synthesized by melting mixtures of the elements in an arc-furnace under an argon atmosphere. Single crystals were obtained by special heat treatment in welded niobium ampoules. The compounds were investigated by means of powder and single crystal X-ray diffraction. All compounds crystallize in the $ThCr_2Si_2$-type structure with space group *I*4/*mmm* (No. 139): $a$ = 3.939(1) Å, $c$ = 10.185(1) Å, $R_1$ = 0.045, 85 $F^2$ values, 8 variable parameters for $CaFe_2Si_2$; $a$ = 4.0612(3) Å, $c$ = 9.9373(9) Å, $R_1$ = 0.030, 90 $F^2$ values, 10 variable parameters for $CaFe_{0.68(6)}Rh_{1.32(6)}Si_2$; $a$ = 4.0695(1) Å, $c$ = 9.9841(3) Å, $R_1$ = 0.031, 114 $F^2$ values, 8 variable parameters for $CaRh_2Si_2$; $a$ = 3.974(1) Å, $c$ = 10.395(1) Å, $R_1$ = 0.036, 95 $F^2$ values, 8 variable parameters for $SrCo_2Si_2$. The structure of $SrCo_2Si_2$ contains isolated $[Co_2Si_2]^{2-}$ 2D-layers in the *ab* plane whereas in $CaFe_{2-x}Rh_xSi_2$ the $[T_2Si_2]$ layers ($T$ = Fe, Rh) are interconnected along the $c$ axis via Si-Si bonds resulting in a three-dimentional (3D) $[T_2Si_2]^{2-}$ and therefore belong to the so-called collapsed form of the $ThCr_2Si_2$-type structure. The $SrCo_2Si_2$ and $CaRh_2Si_2$ are isoelectronic to the parent 122 iron-pnictide superconductors $AeFe_2As_2$, whereas $CaFe_2Si_2$ is a full substituted variant (As/Si) of $CaFe_2As_2$. The crystal chemistry and chemical bonding in the title compounds are discussed in terms of LMTO band structure calculations and a topological analysis using the Electron Localization Function (ELF).

## 1. Introduction

The discovery of the superconducting behavior of $Ba_{0.6}K_{0.4}Fe_2As_2$ [1] renewed the interest in the ternary intermetallic compounds with the composition $AT_2X_2$ ($A$ = alkali/ alkaline earth/ rare earth element; $T$ = transition metal and $X$ = element of the 13-15$^{th}$ group). More than 600 compounds are already known that generally crystallize with the $ThCr_2Si_2$ structure type [2-4]. Interestingly the ternary parent compounds of the Fe-pnictide superconductors show no superconducting behavior but superconductivity emerges under external pressure or by chemical substitution [5]. Recently low temperature onset values ($T_c$) for the superconducting state at ambient pressure have been reported for some $AFe_2X_2$ ($A$ = K, Rb, Cs; $X$ = As [6, 7], Se [8, 9]). In case of 122 iron selenides it was observed that the Fe-vacancies in the crystal lattice play a crucial role in the emergence of superconductivity. Generally, for the occurrence of superconducting behavior under ambient pressure the parent 122 compounds need some kind of substitution at one of the atom sites which can be categorized into electron-poor [1], electron-rich [10] or electron-equal substitution [11]. Due to the large variety of compounds with $ThCr_2Si_2$-type structure they have high chemical flexibility and plenty of substitution possibilities are given. Recently attention was paid also to a series of the iron- and arsenic-free $AeT_2X_2$ phases ($Ae$ = alkaline earth metal, $T$ = Ni, Pd; $X$ = P, Ge) with very low transition temperatures ($Tc$ ~ 0.3 - 3.0 K) [12-15]. Undoped $RT_2Si_2$ ($R$ = La, Y, Th; $T$ = Ir, Pt) superconductors are also known, crystallizing however in $CaBe_2Ge_2$-type structure [16]. Most of the known 122 transition metal silicides have rare earth metals at the $A$ site. Compounds with alkaline earth metals at this site crystallizing in the $ThCr_2Si_2$-type structure are limited [2, 3, 17-23].

In this work four new phases with alkaline earth metals on the $A$-site crystallizing in the $ThCr_2Si_2$-type structure are presented. Moreover, $CaFe_2Si_2$, $CaRh_2Si_2$ and $SrCo_2Si_2$ are the first ternary compounds in the respective $Ae$/$T$/Si ($Ae$ = Ca, Sr; $T$ = Fe, Rh, Co) phase diagrams. $CaFe_2Si_2$ extends the row of $CaT_2Si_2$ compounds containing 3$d$-block transition metals ($T$ = Co-Zn) [3, 18, 19, 21]. $SrCo_2Si_2$ represents the third ternary $SrT_2Si_2$ compound ($T$ = 3$d$-block transition metals) besides $SrCu_2Si_2$ [23] and $SrZn_2Si_2$ [21]. For $SrT_2Si_2$ the isostructural compounds with $T$ = Pd, Ag have been characterized and reported [20, 22]. $CaRh_2Si_2$, the heavier homolog to the $CaCo_2Si_2$ [3], is only the second $CaT_2Si_2$ compound ($T$ = 4$d$-block transition metals) besides $CaPd_2Si_2$ [20]. $BaRh_2Si_2$ [24] has been reported as well but the compound crystallizes in the $BaIr_2Si_2$ structure type.

One interesting feature for the $AT_2X_2$ compounds is the phenomenon of the lattice collapse. Depending on the distance between two $[T_2X_2]^{2-}$ layers, a possible covalent *X*-*X* bond, which leads to the collapse of the structure, can be observed [25, 26]. The nature of the *X*-*X* interactions importantly changes the structural parameters and has a significant influence on the electronic structure. The shape of the $[TX_4]$ tetrahedra and the distance separating the adjacent $[T_2X_2]^{2-}$ layers have also a great influence on the electronic states at the Fermi level. Since $SrCo_2Si_2$ and $CaRh_2Si_2$ are isoelectronic to the parent Fe-pnictide superconductors $AeFe_2As_2$, and $CaFe_2Si_2$ is full substituted variant of $CaFe_2As_2$ [27] at the *X* site, comparison of their electronic bonding situation will be of special interest.

## 2. Experimental Section

### *2.1. Syntheses*

Starting materials for the syntheses of the title compounds were commercially available elements with high purity: ingots of calcium (Alfa Aesar, 99.5 %, redistilled prior to use), lumps of strontium (ChemPur, 98+ %, redistilled prior to use), rhodium shots (Alfa Aesar, 99.9+ %), iron granules (Alfa Aesar, 99.98 %) and silicon pieces (Alfa Aesar, 99.999 %). The sample preparations, arc-melting processes as well as the filling and sealing of the ampoules were performed in an argon-filled glovebox (MBraun 20 G, argon purity 99.996 %). Stoichiometric amounts of the transition metals cobalt, iron or rhodium and silicon (1:1) were pre-melted in an arc-furnace (Mini Arc Melting System, MAM-1, Johanna Otto GmbH).

For the synthesis of $CaFe_2Si_2$ 0.3 g ground FeSi powder was transferred to a niobium ampoule and the alkaline earth metal was added in the ratio 1:2 (Ca:"FeSi", overall mass of 0.372 g), afterwards the ampoule was sealed. The sample was heat-treated in a resistance furnace (HTM Reetz GmbH, Model LOBA 1200-40-600, connected to a thermo controller (EUROTHERM Deutschland GmbH). The ampoule was placed in an evacuated silica tube, heated to approximately 1223 K (Nabertherm, Controller P333) and held at this temperature for 24 h. Afterwards the temperature was reduced to approximately 1073 K in 25 h, kept at this temperature for 240 h and then cooled to room temperature within 4.5 h.

The $CaRh_2Si_2$ sample was prepared in an arc furnace by first pre-melting stoichiometric amounts of Rh and Si (1:1) and then adding Ca ingots in excess (1.5:2 Ca:"RhSi") to counteract the evaporation losses. The sample was afterward annealed in an induction furnace (Hüttinger Elektronik, Freiburg, Typ TIG 2.5/300 equipped with a Sensor

Therm Metis MS06 pyrometer for temperature monitoring) to improve homogeneity and single crystal quality. The sample was ground, pressed to a tablet and sealed in a tantalum ampoule. The ampoule was placed in a water-cooled sample chamber of an induction furnace, heated under a flow of argon to approximately 1273 K and held at this temperature for 1 h. The sample was cooled to room temperature within 2 h.

The quaternary compound $CaFe_{0.68(6)}Rh_{1.32(6)}Si_2$ was prepared by melting the elements in a stoichiometric composition of 1:1:1:2 (Ca:Rh:Fe:Si). The sample was heated in a welded Ta-ampoule to 1273 K for 12 h and then reducing the temperature to 1223 K holding it for 48 h in a resistance furnace. The sample was then ground and pelletized and annealed at 1423 K for 12 h followed by 100 h at 1323 K.

For the $SrCo_2Si_2$ compound (overall mass of 0.5 g) strontium was added to the "CoSi" pellet (Sr : "CoSi" ratio of 1 : 2) and melted three consecutive times in an arc-furnace to obtain higher homogeneity of the product. Single crystals were obtained from that sample. In addition a sample was synthesized in a niobium ampoule containing the pre-melted "CoSi" ground to a fine powder and stoichiometric amount of Sr (Sr : "CoSi" ratio of 1 : 2) and heat treated analog to the $CaFe_2Si_2$ sample.

The samples of $SrCo_2Si_2$ and $CaFe_2Si_2$ are not stable against air and moisture due to SrSi and $Ca_5Si_3$ impurities, respectively. Both Rh-containing samples are air stable. However, all reported intermetallic compounds are stable against air and moisture. Small single crystals of the three compounds could be isolated; they exhibited a plate-like crystal shape.

### *2.2. X-ray investigations*

The purity of the $SrCo_2Si_2$ and $CaRh_2Si_2$ samples was checked at room temperature using a STOE Stadi P powder diffractometer with Cu-$K_\alpha$ radiation ($\lambda$ = 1.54056 Å, Ge(111) monochromator). The iron containing samples were tested with a STOE Stadi P powder diffractometer with Mo-$K_\alpha$ radiation ($\lambda$ = 0.71070 Å, Ge(111) monochromator). The powder X-ray diffraction pattern of the samples showed that all samples contained besides the ternary also known binary compounds as impurities. In case of $SrCo_2Si_2$ the additional phase was air sensitive SrSi and air stable CoSi, for $CaFe_2Si_2$ it was air sensitive $Ca_5Si_3$ and air stable FeSi, for $CaRh_2Si_2$ air stable RhSi was detected as a side phase and for $CaFe_{0.68(6)}Rh_{1.32(6)}Si_2$ air stable FeSi was observed. Attempts to synthesize purer samples have failed. The tetragonal lattice parameters for all three ternary title compounds (see Table 1) were obtained from least-square fits of the powder X-ray diffraction data using WinXPOW [28]. The powder diffraction patterns are given in the Supporting Information, Figure S1-S4.

For the single crystal measurements the crystals were mounted on the tip of a glass fibre under a microscope using nail polish as glue. Single-crystal intensity data for $SrCo_2Si_2$, $CaRh_2Si_2$ and $CaFe_{0.68(6)}Rh_{1.32(6)}Si_2$ were collected at room temperature using an Oxford Diffractions Xcalibur 3 diffractometer with graphite monochromatized Mo-$K_\alpha$ ($\lambda$ = 0.71071 Å) radiation. Single-crystal data for the $CaFe_2Si_2$ compound was collected at room temperature using a Stoe IPDS-2T image plate diffractometer with graphite monochromatized Mo-$K_\alpha$ ($\lambda$ = 0.71071 Å) radiation. The raw data were corrected for background, polarization and Lorentz factor. For $SrCo_2Si_2$ $CaRh_2Si_2$ and $CaFe_{0.68(6)}Rh_{1.32(6)}Si_2$ an empirical absorption correction was applied [29] and for $CaFe_2Si_2$ a semi-empirical absorption correction was performed [30-32].

The starting atomic parameters were deduced from an automatic interpretation of direct methods with SHELXS-97 [33]. For the subsequent refinement for the ternary compounds the lattice parameters determined from the X-ray powder diffraction measurements were given as input data. In case of $CaFe_{0.68(6)}Rh_{1.32(6)}Si_2$ the lattice parameters from single crystal X-ray diffraction measurement were used. The structures were then refined using SHELXL-97 (full-matrix least-square on $F_O^2$) [34] with anisotropic atomic displacement parameters for all atoms. The occupancy parameters for each atom were refined in separate least-squares cycles to check the correct composition of the compounds. The refinements showed that all atom sites in $CaFe_2Si_2$, $CaRh_2Si_2$ and $SrCo_2Si_2$ are fully occupied, whereas in $CaFe_{0.68(6)}Rh_{1.32(6)}Si_2$ the transition metal site was occupied to 66(3) % with Rh and to 34(3) % with Fe. No significant residual peaks were observed for a difference electron-density (Fourier) synthesis. The interatomic distances given in the article refer to the single crystal diffraction data of $SrCo_2Si_2$, $CaFe_2Si_2$ and $CaRh_2Si_2$ using the lattice parameters determined by X-ray powder diffraction as input for the structure refinement. The relevant crystallographic data for the data collection and the structure refinement are listed in Table 1. Further details of the structure determinations may be obtained from: Fachinformationszentrum Karlsruhe, D-76344 Eggenstein-Leopoldshafen, Germany (fax: +49-7247-808-666; email: crysdata@fiz-karlsruhe.de) by quoting the Registry No's. CSD-425467 ($CaFe_2Si_2$), -425468 ($CaFe_{0.68(6)}Rh_{1.32(6)}Si_2$), -425469 ($CaRh_2Si_2$), and -425470 ($SrCo_2Si_2$).

After data collection the single crystals were analyzed by EDX measurements using a JEOL SEM 5900 LV Scanning Electron Microscope equipped with an Oxford Instruments

INCA energy dispersive X-ray microanalysis system. The semi-quantitative EDX analysis reveals the elemental compositions (values given in at. %): Sr 24(3), Co 38(5) and Si 38(5) for $SrCo_2Si_2$; Ca 21(2), Fe 42(4) and Si 38(3) for $CaFe_2Si_2$; Ca 20(2), Rh 42(4) and Si 38(4) for $CaRh_2Si_2$; Ca 19(2), Fe 12(2), Rh 30(4) and Si 39(4) for $CaFe_{0.68(6)}Rh_{1.32(6)}Si_2$, which within standard deviations corresponds to the compositions of the title phases.

### *2.3. Magnetic measurements*

Magnetic measurements were performed using a MPMS XL5 SQUID magnetometer (Quantum Design). All samples were investigated with a zero-field cooled / field-cooled (zfc-fc) measurement and magnetization measurement. The temperature range for the zfc-fc measurements was between 1.8 and 30 K (1.8-15 K for the $CaRh_2Si_2$ sample), for the field-cooled part a field (H) of $1.5 \cdot 10^{-4}$ T was applied. The magnetic behaviour was measured at room temperature up to 5 T. All data were corrected for the holder and for the diamagnetic contribution of the core electrons.

None of the title compounds show superconductive behavior down to 1.8 K. Due to ferromagnetic impurities the samples of $SrCo_2Si_2$, $CaFe_2Si_2$ and $CaFe_{0.68(6)}Rh_{1.32(6)}Si_2$ showed ferromagnetic behaviour in the measured field range. The $CaRh_2Si_2$ sample showed diamagnetic behaviour for fields higher than 0.5 T, whereas in the measured range between 0.01-0.5 T the magnetization has a positive value indicating paramagnetic influences for $CaRh_2Si_2$.

### *2.4. Electronic structure calculations*

The electronic structures for the ternary title compounds were calculated employing a linear muffin-tin orbital (LMTO) method in the atomic sphere approximation (ASA) in the tight-binding (TB) program [35]. The radii of the muffin-tin spheres and empty spheres were determined after Jepsen and Andersen [36]. The Brillouin zone integrations were performed with a 16 × 16 × 32 special *k*-point grid. The basis set of short-ranged [37] atom-centered TB-LMTOs contained *s*, *d* valence functions for Ca and Sr; *s*-*d* valence functions for Fe, Co, and Rh; *s*, *p* valence functions for Si. Ca 4*p* and Sr 5*p* orbitals were included using a downfolding technique.

The analysis of the chemical bonding is based upon theoretical partial and total density of states (DOS) curves, plots of the crystal orbital Hamilton populations (COHPs) [38], band structures with fatbands and contour line diagrams of the Electron Localization Function (ELF) [39]. From the COHP analyses the contribution of the covalent part of a particular interaction to the total bonding energy of the crystal can be obtained. The atomic orbital character is represented as a function of the band width in the fatband analysis.

## 3. Results and Discussions

The four title compounds $CaFe_2Si_2$, $CaFe_{0.68(6)}Rh_{1.32(6)}Si_2$, $CaRh_2Si_2$ and $SrCo_2Si_2$, adopt the $ThCr_2Si_2$-type structure (Figures 1). The $ThCr_2Si_2$-type structure has been described in detail before [25, 40]. Main motifs of this structure are the ${}_{\infty}^{2}[T_2X_2]^{2-}$ layers which contain $[TX_4]$ edge shared tetrahedra in the *ab* plane. The *T* atoms form a square-planar arrangement in the *ab* plane and are tetrahedrally coordinated by Si atoms that are situated in analogy to a chess board alternating above and below the squares of *T* atoms. The distortion from tetrahedral symmetry is influenced by the *T*-*T*, *T*-*X* interactions in the layers and *X*-*X* interactions between the layers. The tunable distance between the *X* atoms of adjacent layers is an important factor in this structure type which determines if the $[T_2X_2]^{2-}$ layers are restricted in the *ab* plane (two-dimensional) or build a 3D (three-dimensional) network. The layers are separated by cations which are located in the cavities created by eight tetrahedra of two adjacent layers. The structure of $SrCo_2Si_2$ contains isolated $[Co_2Si_2]^{2-}$ 2D-layers in the *ab* plane, whereas the [*T*-Si] layers (*T* = Fe, Rh) in $CaFe_{2-x}Rh_xSi_2$ are interconnected along the *c* axis via Si-Si bonds resulting in a $[T_2Si_2]^{2-}$ 3D-network.

Table 5 lists some structural parameters for different $AeT_2Si_2$ (*Ae* = Ca-Ba, *T* = Fe-Zn, Pd, Ag, Rh) [2, 3, 20-23, 41] compounds including the data for the ternary title compounds. In addition the structural parameters for the $AeT_2Ge_2$ (*Ae* = Ca-Ba, T = Mn-Zn) [17, 23] compounds are given in Table S1 (Supporting information).

### 3.1. *Crystal structure of $CaFe_2Si_2$*

$CaFe_2Si_2$ is the first reported compound in the $AeFe_2X_2$ systems (*Ae* = alkaline-earth metal, *X* = group 14 elements). Although the parent compounds of the iron pnictides superconductors $AeFe_2Pn_2$ have been studied extensively, the analogue tetrelides have not been investigated so far. The substitution of Co in $CaFe_2Si_2$ does almost not influence the lattice parameter *a* (3.939(1) Å for $CaFe_2Si_2$ and 3.92 Å for $CaCo_2Si_2$ [3]) and the interatomic distances $d_{T\text{-}T}$ in the $[T_2X_2]^{2-}$ layers are very similar (2.785(1) Å and 2.77 Å for $CaFe_2Si_2$ and $CaCo_2Si_2$, respectively). However it has a significant impact on the lattice parameter *c*, heights of Si atoms above *T* layer, the distances $d_{T\text{-Si}}$, $d_{\text{Si-Si}}$ and the tetrahedral angle $\alpha$. The *c* parameter is almost 0.2 Å longer in the Fe-compound (10.185(1) Å vs 9.92 Å for $CaCo_2Si_2$) and correspondingly the volume of the unit cell decreases from 158.0(1) Å$^3$ in $CaFe_2Si_2$ to 152.1 Å$^3$ for $CaCo_2Si_2$. The $[T_2X_2]^{2-}$ layers are further apart due to the increase of the *c* parameter. The heights of Si atoms above *T* layer are 1.269 Å for $CaFe_2Si_2$ and 1.190 Å for $CaCo_2Si_2$, which is also reflected in the interlayer $d_{\text{Si-Si}}$ (2.555(9) Å in $CaFe_2Si_2$ and 2.58 Å in $CaCo_2Si_2$). Hoffmann and Zheng have observed the trend that the $d_{\text{X-X}}$ values should decrease for a system $AeT_2X_2$ (same *Ae* and *X*) when moving from the left-hand side of the periodic table to the right-hand side for the *T* element [26]. They placed their observation on ternary compounds where *T* = P. The P-containing compounds belong to the compounds with uncollapsed $ThCr_2Si_2$ type structure for transition metals on the left-hand side of the periodic table. In the case of $CaFe_2Si_2$ and $CaCo_2Si_2$ a differentiation of this trend is observed since the Fe-compound has the shorter $d_{\text{Si-Si}}$ value. Comparing the $ThT_2Si_2$ (*T* = Cr-Cu) crystallographic data a similar effect can be seen when comparing the $ThCo_2Si_2$ compound with the $ThFe_2Si_2$ compound [42]. It seems that the trend of decreasing $d_{X\text{-}X}$ interactions for $AT_2X_2$ compounds (*A* = alkaline earth metals or rare earth metals, *X* = tetrels), when moving to the left-hand side of the Periodic Table (for *T* = Mn-Zn), is not a uniform trend for compounds with collapsed $ThCr_2Si_2$ structure type.

The geometry of the tetrahedra in $[T_2Si_2]^{2-}$ changes as a result of the different height of Si atoms above *T* layer. In $CaFe_2Si_2$ the $d_{\text{Fe-Si}}$ value is slightly elongated (2.343(2) Å) compared to $d_{\text{Co-Si}}$ of 2.29 Å in $CaCo_2Si_2$ and the $\alpha$ values decreases (114.4(2) ° for $CaCo_2Si_2$ and 117.4 ° for $CaCo_2Si_2$).

It should be mentioned that the structure of the well-known superconductor compound $CaFe_2As_2$ [27], where Si is fully substituted (Si/As), contains no *X-X* interlayer bonds at ambient pressure whereas in the high-pressure modification this covalent bonds emerges [43, 44]. $CaFe_2As_2$ has a slightly shorter lattice parameter *a* of 3.87 Å than the $CaFe_2Si_2$ compound (3.939(1) Å), indicating stronger Fe-Fe interactions ($d_{T\text{-}T}$ of 2.74 Å for $CaFe_2As_2$ and 2.785(1) Å for $CaFe_2Si_2$). The values of the lattice parameters *c* and *V* differ noticeably ($c$ = 11.73 Å, $V$ = 175.94 Å$^3$ for ucT-$CaFe_2As_2$ at ambient pressure) [27]. The discrepancy is due to the non-existing As-As bonds along the *c* direction. The height of As atoms above Fe-layer is influenced by the missing As-As bonds (1.367 Å for $CaFe_2As_2$ and 1.269 Å for $CaFe_2Si_2$). The Si-Si-bond in $CaFe_2Si_2$ results in a vertical displacement of the Si atomic *z*-coordinate towards higher values to accommodate the covalent bond compared to the 2-D layered structure of $CaFe_2As_2$. The tetrahedral angle $\alpha$ is close to the ideal tetrahedral angle of 109.4 ° in ucT-$CaFe_2As_2$ (109.5 °) but is larger in $CaFe_2Si_2$ (114.4(2) °), which is in accordance with the structural properties of other reported 122 arsenides and silicides [3].

### 3.2. *Crystal structure of $CaRh_2Si_2$*

$CaRh_2Si_2$ is isoelectronic to $CaFe_2As_2$ and is analogue to $CaCo_2Si_2$ [3] with the heavier homologue on the transition metal position. The increased covalent radius of Rh (1.350 Å to 1.240 Å for Co) [33, 34] mainly affects the parameters and distances connected to the square transition metal plane in the *ab* plane. The lattice parameter *a* is 0.15 Å larger for $CaRh_2Si_2$ (4.070(1) Å) compared to $CaCo_2Si_2$ (3.92 Å) resulting in an elongation of the distances between the transition metal atoms (2.878(1) Å and 2.77 Å, respectively) and the Si atoms in that plane.

The values for the *c* parameter vary less (9.984(1) Å and 9.92 Å respectively) and so should the distance between the $[T_2X_2]^{2-}$ layers. The heights of Si atoms above *T* layer is larger in the $CaRh_2Si_2$ (1.231 Å) than in $CaCo_2Si_2$ (1.190 Å). The distance between the interlayer Si-Si atoms is shorter ($d_{\text{Si-Si}}$ = 2.529(8) Å for $CaRh_2Si_2$ and 2.58 Å for $CaCo_2Si_2$) resulting in stronger Si-Si interactions. Due to the longer *T*-*T* distance for the Rh-compound and the larger height of Si atoms the *T*-Si distance has to increase (2.378(1) Å for $CaRh_2Si_2$ and 2.29 Å for $CaCo_2Si_2$). Those changes result in almost unchanged tetrahedral angle in the $[T_2X_2]^{2-}$ layers (117.6(3) ° for $CaRh_2Si_2$ and 117.4 ° for $CaCo_2Si_2$). Similar observations can be made for the comparison between $Ae$$Ni_2Si_2$ and $Ae$$Pd_2Si_2$, containing transition metal homologues (see Table 5).

### 3.3. *Crystal structure of $CaFe_{0.68(6)}Rh_{1.32(6)}Si_2$*

The values of the *a* lattice parameter, volume, *T*-*T* and *T*-Si interatomic distances of quaternary compound $CaFe_{0.68(6)}Rh_{1.32(6)}Si_2$ are between those of the side compounds $CaFe_2Si_2$ and $CaRh_2Si_2$ but closer to those of $CaRh_2Si_2$ due to the higher Rh-content, approaching Vegard's law (Tables 1 and 3). However, an anomalous compression of the *c* parameter of $CaFe_{0.68(6)}Rh_{1.32(6)}Si_2$ in comparison to the side ternary compounds is observed. In order to maintain the interlayer Si-Si distance almost the same as in $CaFe_2Si_2$ (with largest *c*) the height of Si atoms above *T* layer is smaller and the tetrahedral angle $\alpha$ is flattened.

This is in contrast to the $CaFe_{2-x}Rh_xAs_2$ solid solution for which nearly linear variety in the lattice parameters was reported, however, only for the region of x = 0 - 0.6 [45]. Nevertheless, when comparing the lattice parameters with those of $CaRh_2As_2$ [46] the deviation from linearity becomes noticeable. Therefore further studies of the *T*/*T* substitution in $AeT_2X_2$ phases are required for better understanding of the structural properties in these systems.

### 3.4. Crystal structure of $SrCo_2Si_2$

$SrCo_2Si_2$ has the smallest lattice parameter *a* and volume ($a = 3.974(1)$ Å, $V = 164.2(1)$ Å$^3$) compared to $SrCu_2Si_2$ and $SrZn_2Si_2$ (4.20 Å and 4.33 Å, 179.4 Å$^3$ and 194.1 Å$^3$, respectively), but the largest *c* parameter (10.395(2) Å, 10.00 Å and 10.35 Å, respectively) [21, 23]. To account for this effect the interaction in the *ab* plane has to be stronger for $SrCo_2Si_2$ but are weakened along the *c* axis, which is reflected in the *T*-*T*, *T*-Si and Si-Si interatomic distances. The elongation of the *c* axis is a result of the different bonding situation between the adjacent tetrahedral-layers. In $SrCu_2Si_2$ the $[T_2X_2]^{2-}$ layers are connected via Si-Si covalent bonds (2.42 Å) resulting in the collapsed $ThCr_2Si_2$ type structure, whereas in $SrCo_2Si_2$ the Si-Si distance is elongated (2.898(5) Å) which indicates that the layers are well separated without covalent bonds in between. This observation is in accordance with the results published by Hoffmann and Zheng [26], stating that $d_{X\text{-}X}$ decreases as the transition metal *T* moves from the left-hand side to the right-hand side in the Periodic Table. At this point the $SrT_2Si_2$ and $SrT_2Ge_2$ systems (*T* = Mn-Zn) show a deviation (Table 5; Supporting information, Table S1). Even though the $d_{\text{Ge-Ge}}$ decreases when going from Co ($d_{\text{Ge-Ge}} = 2.90$ Å) to Ni ($d_{\text{Ge-Ge}} = 2.83$ Å) they are certainly longer than that of a Ge-Ge covalent single bond in diamond-like Ge (2.45 Å). Only the $SrMn_2Ge_2$ ($d_{\text{Ge-Ge}} = 2.63$ Å), possibly $SrCu_2Ge_2$ ($d_{\text{Ge-Ge}} = 2.70$ Å) and definitely $SrZn_2Ge_2$ ($d_{\text{Ge-Ge}} = 2.50$ Å) can be considered as having a covalent interlayer Ge-Ge bond. Since the *X*-*X* interactions are weakend the *T*-*T* and *T*-*X* interactions in the $[T_2X_2]^{2-}$ layer are more prominent as can be seen from the corresponding values in Tables 5 and S1 (Supporting information). For $SrCo_2Si_2$ the *T*-*T* and *T*-Si distances in the $[T_2Si_2]^{2-}$ layers are shorter than for $SrCu_2Si_2$ (2.810(1) Å and 2.296(3) Å compared to 2.97 Å and 2.47 Å, respectively).

Due to the shorter distances in the $[T_2Si_2]^{2-}$ layer and the smaller height of Si atoms above *T* layer (1.150 Å for $SrCo_2Si_2$ and 1.290 Å for $SrCu_2Si_2$ [23]) the tetrahedra are flattened in $SrCo_2Si_2$ resulting in a larger tetrahedra angle α compared to $SrCu_2Si_2$ (119.1(2) ° to 116.9 °). The influence of the deformation of the $[TX_4]$ tetrahedra from an ideal tetrahedron $\alpha = 109.4$ ° has been already discussed for the $CaT_2Ge_2$ compounds [21]. By varying the transition metal from Mn to Zn the tetrahedral angle increases from 111.3 ° for Mn to 119.1 ° for Ni and then decreases to 114.9 ° for Cu and 109.9 ° for Zn (Supporting information, Table S1). The almost ideal tetrahedral structure for $CaZn_2Ge_2$ can be attributed to the filled 3d-shell. Unfortunately, it is impossible to detect a trend for the $SrT_2Si_2$ compounds since only crystallographic data for $T$ = Co, Cu is available.

When comparing $SrCo_2Si_2$ with $CaCo_2Si_2$ the expected differences in the parameters are observed. First of all are the lattice parameters (*a* and *c* parameter and correspondingly *V*) all larger for the Sr-compound due to the large ionic radius of $Sr^{2+}$ (1.18 Å compared to 1.00 Å for $Ca^{2+}$) [47] as are the $d_{\text{Si-Si}}$ values (2.898(5) Å for $SrCo_2Si_2$ and 2.58 Å for $CaCo_2Si_2$) [3]. The heights of Si atoms above Co layer are smaller in $SrCo_2Si_2$ compared to $CaCo_2Si_2$ (1.150 Å and 1.190 Å, respectively). The $CaCo_2Si_2$ compound contains covalent Si-Si bonds along the *c* direction and belongs to the group of compounds with collapsed $ThCr_2Si_2$-type structure. Analogous observations can be made for the corresponding Ge-compounds (Supporting information, Table S1). The value for $d_{T\text{-Si}}$ is very similar (2.29 Å in $CaCo_2Si_2$ and 2.296(3) Å in $SrCo_2Si_2$) but the $d_{T\text{-}T}$ value differentiates slightly (2.77 Å in $CaCo_2Si_2$ and 2.810(1) Å in $SrCo_2Si_2$), which in combination with the lattice parameter and the value of height of Si atoms leads to the higher $\alpha$ value of tetrahedral angle in $SrCo_2Si_2$ (119.9(2) ° *vs* 117.4 ° in $CaCo_2Si_2$). It is necessary to note, that the $SrCo_2Si_2$ is isoelectronic to the parent 122 iron-pnictide superconductors $AeFe_2As_2$.

### *3.5. Electronic structure calculations*

In the title silicides $CaFe_2Si_2$, $CaRh_2Si_2$ and $SrCo_2Si_2$, similar to other $AeT_2X_2$ phases, the bonding between $Ae^{2+}$ and the $[T_2X_2]^{2-}$ layer is mainly ionic, while in the $[T_2X_2]^{2-}$ layers the covalent *T-X* bonds as well as weak metal-metal *T-T* bonds occur. In order to analyse the electronic properties of these compounds the total Density Of States (DOS) as well as the partial DOS were calculated (Figure 2).

Comparison of the DOS profiles for isostructural $CaFe_2Si_2$, $CaRh_2Si_2$ and $SrCo_2Si_2$ allows us to understand the main differences of their electronic structure. No band gap is observed at the Fermi level in all cases indicating metallic properties of the title compounds. The DOS can be divided into two parts separated by a gap: one block at lower energies (from −10.5 to −6.5 eV for $CaFe_2Si_2$, from −11 to −7.5 eV for $CaRh_2Si_2$ and from −11 to −7 eV for $SrCo_2Si_2$) and one from about −5.5 eV ($CaFe_2Si_2$), −6.5 eV ($CaRh_2Si_2$) and −5.5 eV ($SrCo_2Si_2$) up to energies above the Fermi level. The Si-s states are located at the bottom of the valence band (first block). The first half of the second block (from −5 to −1 eV for $CaFe_2Si_2$, from −6.5 to −3.5 eV for $CaRh_2Si_2$ and from −5.5 to −2 eV for $SrCo_2Si_2$) is mainly build up from *T*-d and Si-p partial DOS of similar shape, indicating the hybridisation of these orbitals and correspondingly covalent *T*-Si bonds within $[T_2X_2]^{2-}$ layers. Calculation of the Crystal Overlap Hamiltonian Population (COHP) for this part of the DOS reveals good *T*-Si interactions (Table 3, Supporting Information, Figures S5-7). The upper half of the second block (from −1 eV for $CaFe_2Si_2$, −3.5 eV for $CaRh_2Si_2$ and −2 eV for $SrCo_2Si_2$ to $E_F$) is dominated by the *T*-d orbitals. It is well established that the electronic bands around the Fermi level are responsible for superconductivity. These bands are formed mainly by the d-states of the *T* atoms for $CaFe_2Si_2$ and, to a lesser effect, for $SrCo_2Si_2$ whereas the pseudogap is now situated at $E_F$ for $CaRh_2Si_2$ with smaller *T*-d state contributions. The contributions of Ca/Sr states to the valence bands are negligible since these atoms are in the form of cations $Ae^{2+}$ in the $AeT_2X_2$ structures and serve as electron-donors. The contribution of the Si atoms to the Fermi surface is small but non-zero.

The density of states of the examined silicides, going from $CaFe_2Si_2$ to $CaRh_2Si_2$ and $SrCo_2Si_2$, clearly shows a shifting of the Fermi level to conduction band in accordance with increasing of electron count. A local maximum at the Fermi level in $CaFe_2Si_2$, formed predominantly by Fe-d states, can be correlated to a degree of structural or magnetic instability observed in Fe-containing 122 superconductors. The rhodium silicide $CaRh_2Si_2$ has an increased valence electron count (+2*e*) as compared to $CaFe_2Si_2$, resulting in a shift of $E_F$ from the local maximum DOS in $CaFe_2Si_2$ to a pseudogap between two local maxima in $CaRh_2Si_2$. Similar trend is observed for homologue $CaCo_2Si_2$ [3] (Supporting information, Figure S8), where the pseudogap is observed near the Fermi level. From $CaRh_2Si_2$ to $SrCo_2Si_2$ further shifting of $E_F$ followed by the lowering of band dispersion are observed. Despite the same electron count in $CaRh_2Si_2$ and $SrCo_2Si_2$, the bonding situation in these structures seems to be different. The non-bonded Si atoms ($d_{\text{Si-Si}}$ = 2.898(5) Å) in $SrCo_2Si_2$ have the formal electron count $(2\times Si)^{8-}$, whereas the singly bonded Si-Si dimer in $CaRh_2Si_2$ has a formal count of electrons $(Si\text{-}Si)^{6-}$, like the two non-bonded $(2\times As)^{6-}$ in uncollapsed-tetragonal (ucT) $Sr(Ba)Rh_2As_2$ [48, 49]. This in turn affects the electron configuration of transition metal *T* atoms, taking into account that the charge of the cations *Ae* remains the same, and results in the shift of DOS near Fermi level. In this regard, the collapsed tetragonal (cT) $CaFe_2Si_2$ with $(Si\text{-}Si)^{6-}$ dimer is electronically close to the uncollapsed-tetragonal (ucT) $CaFe_2As_2$ with non-bonded $(2\times As)^{6-}$, having similar formal charge distributions: $Ca^{2+}(Fe^{2+})_2(Si\text{-}Si)^{6-}$ *vs* $Ca^{2+}(Fe^{2+})_2(As^{3-})_2$. This is reflected in the similarities of the corresponding DOS: for $CaFe_2As_2$ a similarly shaped density of states of Fe *d* bands near $E_F$ and a pseudogap have also been observed [50].

The situation of the near $E_F$ region in the DOS of the examined silicides is mirrored in the fatband analyses (Figures S9-S11, Supporting information). The contribution of the Fe $3d_{x^2-y^2}$ (pointing directly towards the nearest neighbor Fe atoms) and $3d_{xz,yz}$ orbitals is largest close to $E_F$ for $CaFe_2Si_2$, which is indicated by the sharp peaks in the density of states. These orbitals are moved to lower energies in $CaRh_2Si_2$ and $SrCo_2Si_2$. The conduction band of the studied compounds has mainly $T$-$d_{xy}$ character, which lies near to $E_F$ in $SrCo_2Si_2$, indicated by the sharp peak in DOS. The $T$-$d_{xz,yz}$ orbitals, pointing directly toward Si atoms, show strong hybridization in their lower part of energy with Si-*p* orbitals, indicating strong covalent *T*-Si bonding. However, several Si-*p* bands cross $E_F$ in the mixing with *T-d* orbitals. In $CaFe_2Si_2$ the Si-*p* with Fe-$d_{z^2}$ bands cross $E_F$ mostly in the section Z → Γ (crystallographic *c* direction), whereas in $CaRh_2Si_2$ the Si-*p* orbitals crossing $E_F$ are hybridized along Z → Γ with Rh-$d_{z^2,xz,yz}$ and along Γ → X, P → N (crystallographic *ab*-plane) with Rh-$d_{xz,yz}$ orbitals. In $SrCo_2Si_2$ the Si-*p* bands that cross $E_F$ are hybridized with Co-$d_{z^2}$ along Z → Γ and with Co-$d_{xz,yz,x^2-y^2}$ along Γ → X, P → N. This underlines the influence of *T*-Si bonding interactions on the Fermi surface.

Further insight in the nature of the bonding situation can be provided analysing the electron density in real space using the Electron Localization Function (ELF), sketched in Figure 3. The ELF reveals clearly a bisynaptic valence basin (> 0.72) between the Si atoms for the $CaFe_2Si_2$ and $CaRh_2Si_2$ structures. The basins are orientated towards each other indicating a localized covalent bond. Similar interlayer *X-X* bonds have been described in $CaCo_2Si_2$ and $CaNi_2Ge_2$ [3, 51]. Due to the longer Si-Si distance the ELF reveals no bisynaptic valence basin for $SrCo_2Si_2$ but non-bonding (monotactic) ELF domains (free electron pairs) which are located at the Si atoms with their orientation towards each other. These pairs are arranged closely and touch each other, indicating only very weak bonding interactions between the Si atoms (iCOHP value of 0.93 eV). In $CaFe_2Si_2$, $CaRh_2Si_2$, and $SrCo_2Si_2$ the ELF attractors at low values (0.39, 0.29 and 0.39, respectively) were observed between *T* atoms, forming a planar square net indicating directed bonding interactions. This is in agreement with results obtained for $LaT_2Ge_2$ ($T$ = Mn, Fe, Co) [52].

## 4. Conclusions

The title intermetallic compounds $CaFe_2Si_2$, $CaFe_{0.68(6)}Rh_{1.32(6)}Si_2$, $CaRh_2Si_2$ and $SrCo_2Si_2$ crystallise in the $ThCr_2Si_2$-type structure and show the tunable Si-Si distance.

$CaFe_2Si_2$, $CaRh_2Si_2$ and $SrCo_2Si_2$ are the first ternary compounds in the respective *Ae*/*T*/Si (*Ae* = Ca, Sr; *T* = Fe, Rh, Co) phase diagrams. The anionic substructure in $CaFe_{2-x}Rh_xSi_2$ (x = 0, 1.32, 2) is best described as a three-dimensional ${}^3_\infty[T_2Si_2]^{2-}$ network with covalent Si-Si bonds between ${}^2_\infty[T_2Si_2]^{2-}$ layers and distances almost independent of *T* (2.529(8) - 2.555(9) Å). $SrCo_2Si_2$ consists of discrete two-dimensional ${}^2_\infty[Co_2Si_2]^{2-}$ layers ($d_{Si-Si}$ = 2.898(5) Å). The comparison with other $AeT_2Si_2$ compounds (*T* = Co-Zn) reveals that the change of the transition metal from the left-hand side to the right-hand side in the Periodic Table doesn't induce a linear decrease of the *X*-*X* distance, as was observed for the isostructural phosphides [26]. Analogously, no linear changes of geometrical parameters can be observed during the Fe/Rh substitution in $CaFe_{2-x}Rh_xAs_2$ solid solution. This can be explained by complex interplay between covalent, metallic and ionic interactions in $ThCr_2Si_2$-type compounds. The major contribution to the structural changes traces back to the size of the alkaline earth metal. The variation from the smaller calcium atom to the larger homologues strontium or barium in $AeT_2X_2$ structures induces not only an increase of the *T*-*T* and *X*-*X* distances but also a flattening of the tetrahedral angle *X*-*T*-*X* (Table 5, S1). Interestingly, the series of electron richer compounds $RFe_2Si_2$, $RRh_2Si_2$ and $RCo_2Si_2$ (*R* – rare-earth metal) [3] are known with $R^{3+}$cations in the cavities of $[T_2Si_2]^{3-}$ network, underlying the high chemical and electronic flexibility of $ThCr_2Si_2$-type structure. The results of the quantum chemical calculations concerning DOS, COHP and the band structures including fatbands have been discussed. The topological analysis of the ELF depicts clearly the Si-Si bond in $CaFe_2Si_2$ and $CaRh_2Si_2$ and the non-bonding Si-Si in $SrCo_2Si_2$, as well as weak directed bonding interactions between the transition metal atoms.

**Acknowledgments**

This research was supported by the Deutsche Forschungsgemeinschaft (Priority Program 1458).

**Appendix A. Supplementary Information**

Supplementary data associated with this article can be found in the online version at…

**Table 1** Crystal Data and structure refinement for $CaFe_2Si_2$, $CaFe_{0.68(6)}Rh_{1.32(6)}Si_2$, $CaRh_2Si_2$ and $SrCo_2Si_2$. All data given in this table refer to X-ray single crystal data at T = 293 K refined with cell parameters taken from X-ray powder data for ternary compounds.

| **Empirical formula** | **$CaFe_2Si_2$** | **$CaFe_{0.68(6)}Rh_{1.32(6)}Si_2$** | **$CaRh_2Si_2$** | **$SrCo_2Si_2$** |
|---|---|---|---|---|
| Formula weight / g·mol$^{-1}$ | 207.96 | 264.43 | 302.08 | 261.66 |
| Space group, Z | *I*4/*mmm*, 2 | *I*4/*mmm*, 2 | *I*4/*mmm*, 2 | *I*4/*mmm*, 2 |
| Unit cell dimensions / Å | $a$ = 3.939(1) | $a$ = 4.059(1) | $a$ = 4.070(1) | $a$ = 3.974(1) |
| (powder data; single crystal data for $CaFe_{0.68(6)}Rh_{1.32(6)}Si_2$) | $c$ = 10.185(1) | $c$ = 9.939(1) | $c$ = 9.984(1) | $c$ = 10.395(2) |
| | $V$ = 158.0(1) | $V$ = 163.8(1) | $V$ = 165.3(1) | $V$ = 164.2(1) |
| Calculated density / g·cm$^{-3}$ | 4.372 | 5.363 | 6.068 | 5.293 |
| Absorption coefficient / mm$^{-1}$ | 11.254 | 11.591 | 11.963 | 26.607 |
| $F$(000) | 200 | 246 | 276 | 240 |
| Crystal size / mm | 0.04 x 0.04 x 0.005 | 0.12 x 0.06 x 0.02 | 0.05 x 0.04 x 0.01 | 0.07 x 0.06 x 0.01 |
| θ range / ° | 4.00 to 29.15 | 4.10 to 29.70 | 4.08 to 32.79 | 3.91 to 32.49 |
| Range in hkl | ± 5, ± 5, ±13 | −5≤ k ≤ 3, ± 5, ± 13 | ± 6, ± 5, −14 ≤ l ≤ 15 | ± 5, −4 ≤ k ≤ 5, ± 14 |
| Reflections collected | 1463 | 1335 | 1570 | 1418 |
| Independent reflections | 85 ($R_{int}$ = 0.1107) | 90 ($R_{int}$ = 0.0368) | 114 ($R_{int}$ = 0.0782) | 95 ($R_{int}$ = 0.097) |
| Reflections with $I \geq 2\sigma(I)$ | 68 ($R_\sigma$ = 0.0318) | 89 ($R_\sigma$ = 0.0115) | 90 ($R_\sigma$ = 0.0301) | 87 ($R_\sigma$ = 0.033) |
| Data/parameters | 68/8 | 90/10 | 90/9 | 87/8 |
| GOF on $F^2$ | 1.317 | 1.763 | 1.093 | 1.163 |
| Final $R$ indices [$I \geq 2\sigma(I)$] | $R_1$ = 0.0452 | $R_1$ = 0.030 | $R_1$ = 0.026 | $R_1$ = 0.036 |
| | w$R_2$ = 0.103 | w$R_2$ = 0.062 | w$R_2$ = 0.073 | w$R_2$ = 0.090 |
| $R$ indices (all data) | $R_1$ = 0.068 | $R_1$ = 0.030 | $R_1$ = 0.035 | $R_1$ = 0.038 |
| | w$R_2$ = 0.112 | w$R_2$ = 0.063 | w$R_2$ = 0.075 | w$R_2$ = 0.091 |
| Largest diff. peak and hole / e·Å$^{-3}$ | 0.926 and −0.850 | 1.234 and −0.808 | 3.943 and −2.518 | 1.526 and −1.320 |

**Table 2** Atomic coordinates and isotropic equivalent displacement parameters /Å$^2$ × 10$^3$ for $CaFe_2Si_2$, $CaFe_{0.68(6)}Rh_{1.32(6)}Si_2$, $CaRh_2Si_2$ and $SrCo_2Si_2$ (space group *I*4/*mmm*, Z = 2).

| **Atom** | **Wyckoff position** | ***x*** | ***y*** | ***z*** | $U_{eq}$ **/ Å$^2$ x 10$^3$** |
|---|---|---|---|---|---|
| **$CaFe_2Si_2$** | | | | | |
| Ca | 2*a* | 0 | 0 | 0 | 19(1) |
| Fe | 4*d* | 0 | 1/2 | 1/4 | 17(1) |
| Si | 4*e* | 0 | 0 | 0.3746(4) | 18(2) |
| **$CaFe_{0.68(6)}Rh_{1.32(6)}Si_2$** | | | | | |
| Ca | 2*a* | 0 | 0 | 0 | 8(1) |
| Fe/Rh | 4*d* | 0 | 1/2 | 1/4 | 5(1) |
| Si | 4*e* | 0 | 0 | 0.3716(4) | 7(1) |
| **$CaRh_2Si_2$** | | | | | |
| Ca | 2*a* | 0 | 0 | 0 | 12(1) |
| Rh | 4*d* | 0 | 1/2 | 1/4 | 9(1) |
| Si | 4*e* | 0 | 0 | 0.3733(4) | 12(1) |
| **$SrCo_2Si_2$** | | | | | |
| Sr | 2*a* | 0 | 0 | 0 | 12(1) |
| Co | 4*d* | 0 | 1/2 | 1/4 | 11(1) |
| Si | 4*e* | 0 | 0 | 0.3606(3) | 12(1) |

**Table 3** Interatomic distances calculated with the lattice parameters taken from X-ray powder data for $SrCo_2Si_2$, $CaFe_2Si_2$, $CaRh_2Si_2$ and from single crystal data for $CaFe_{0.68(6)}Rh_{1.32(6)}Si_2$ (space group *I*4/*mmm*, Z = 2) and selected corresponding integrated crystal orbital Hamilton populations (-iCOHPs) values at $E_F$.

| | | **distance / Å** | **iCOHP / eV** | | | **distance / Å** | **iCOHP / eV** |
|---|---|---|---|---|---|---|---|
| **$CaFe_2Si_2$** | | | | | | | |
| Ca | -Ca | 3.938(1) | - | Si | -Fe | 2.343(2) | 2.71 |
| | -Fe | 3.219(1) | - | | -Si | 2.555(9) | 1.98 |
| | -Si | 3.064(2) | 0.65 | Fe | -Fe | 2.785(1) | 1.17 |
| **$CaFe_{0.68(6)}Rh_{1.32(6)}Si_2$** | | | | | | | |
| Ca | -Ca | 4.059(1) | - | Si | -Fe/Rh | 2.362(2) | - |
| | -Fe/Rh | 3.208(1) | - | | -Si | 2.552(6) | - |
| | -Si | 3.141(2) | - | Fe/Rh | -Fe/Rh | 2.870(1) | - |
| **$CaRh_2Si_2$** | | | | | | | |
| Ca | -Ca | 4.069(2) | - | Si | -Rh | 2.378(1) | 2.68 |
| | -Rh | 3.220(1) | - | | -Si | 2.529(8) | 2.10 |
| | -Si | 3.143(2) | 0.65 | Rh | -Rh | 2.878(1) | 1.02 |
| **$SrCo_2Si_2$** | | | | | | | |
| Sr | -Sr | 3.974(1) | - | Si | -Co | 2.296(3) | 2.80 |
| | -Co | 3.271(1) | - | | -Si | 2.898(5) | 0.93 |
| | -Si | 3.162(3) | 0.66 | Co | -Co | 2.810(1) | 0.94 |

**Table 4** Tetrahedral angles $\alpha$ / ° within the [*T*-Si] layer of the title compounds.

| **Compound** | **$\alpha$ (Si-T-Si) / °** |
|---|---|
| $CaFe_2Si_2$ | 114.4(2) |
| $CaFe_{0.68(6)}Rh_{1.32(6)}Si_2$ | 118.5(8) |
| $CaRh_2Si_2$ | 117.6(3) |
| $SrCo_2Si_2$ | 119.9(2) |

**Table 5** Lattice parameters, atomic distances and tetrahedral angles ($\alpha$) of $AeT_2Si_2$ ($Ae$ = Ca, Sr, Ba; $T$ = Fe, Co, Ni, Cu, Zn, Pd, Ag) compounds ($I4/mmm$ space group, Z = 2).

| **Compound** | ***a* / Å** | ***c* / Å** | ***V* / Å$^3$** | ***c*/*a*** | ***d*(*T*-*T*) / Å** | ***d*(*T*-Si) / Å** | ***d*(Si-Si) / Å** | **$\alpha$ / °** | **Ref.** |
|---|---|---|---|---|---|---|---|---|---|
| $CaFe_2Si_2$ | 3.939(1) | 10.185(1) | 158.0(1) | 2.59 | 2.785(1) | 2.343(2) | 2.555(9) | 114.4(2) | * |
| $CaCo_2Si_2$ | 3.92 | 9.92 | 152.1 | 2.53 | 2.77 | 2.29 | 2.58 | 117.4 | [3] |
| $CaNi_2Si_2$ | 3.99 | 9.67 | 153.8 | 2.43 | 2.82 | 2.31 | 2.49 | 119.1 | [41] |
| $CaCu_2Si_2$ | 4.04 | 10.0 | 163.22 | 2.48 | 2.86 | 2.42 | 2.32 | 112.9 | [23] |
| $CaZn_2Si_2$ | 4.17 | 10.58 | 184.2 | 2.54 | 2.95 | 2.55 | 2.36 | 109.9 | [21] |
| $SrCo_2Si_2$ | 3.974(1) | 10.395(2) | 164.2(1) | 2.62 | 2.810(1) | 2.296(3) | 2.898(5) | 119.9(2) | * |
| $SrCu_2Si_2$ | 4.20 | 10.00 | 176.4 | 2.48 | 2.97 | 2.47 | 2.42 | 116.9 | [23] |
| $SrZn_2Si_2$ | 4.33 | 10.35 | 194.1 | - | - | - | - | - | [21]# |
| $BaZn_2Si_2$ | 4.50 | 10.20 | 206.4 | 2.27 | 3.18 | 2.61 | 2.43 | 118.6 | [2] |
| $CaRh_2Si_2$ | 4.070(1) | 9.984(1) | 165.3(1) | 2.45 | 2.878(1) | 2.378(1) | 2.529(8) | 117.6(3) | * |
| $CaPd_2Si_2$ | 4.22 | 9.77 | 173.7 | 2.32 | 2.98 | 2.43 | 2.44 | 119.9 | [20] |
| $SrPd_2Si_2$ | 4.31 | 9.88 | 183.5 | 2.29 | 3.05 | 2.48 | 2.47 | 120.4 | [20] |
| $SrAg_2Si_2$ | 4.38 | 10.48 | 201.1 | 2.39 | 3.10 | 2.63 | 2.33 | 112.7 | [22] |

* this work

# Cell parameters from X-ray powder data, no single crystal data measured to confirm $ThCr_2Si_2$ type structure.

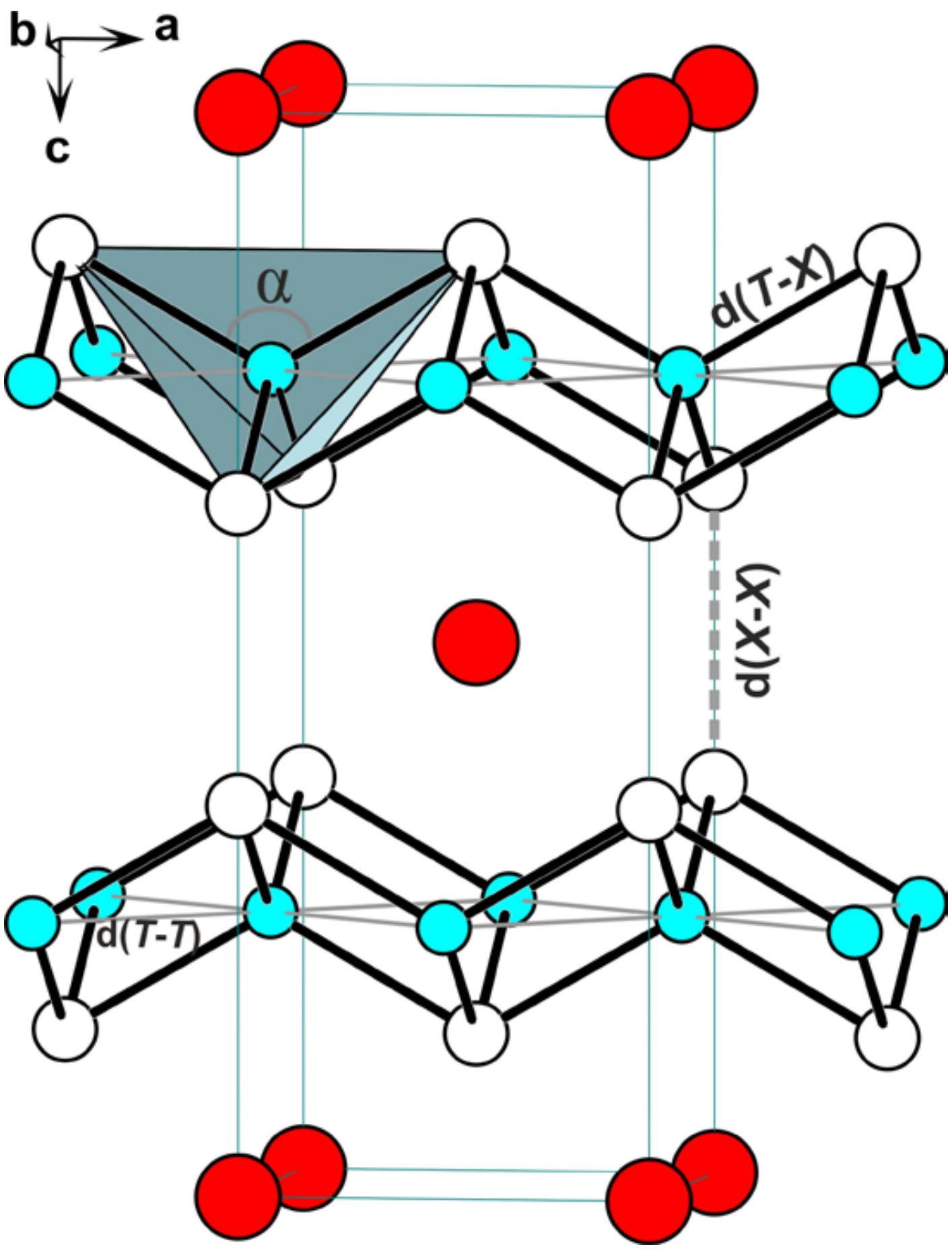

**Figure 1**. Structure of $SrCo_2Si_2$ (Sr - red, Co - blue, Si - white) displayed as a model including selected structural parameters for the $ThCr_2Si_2$-type structure.

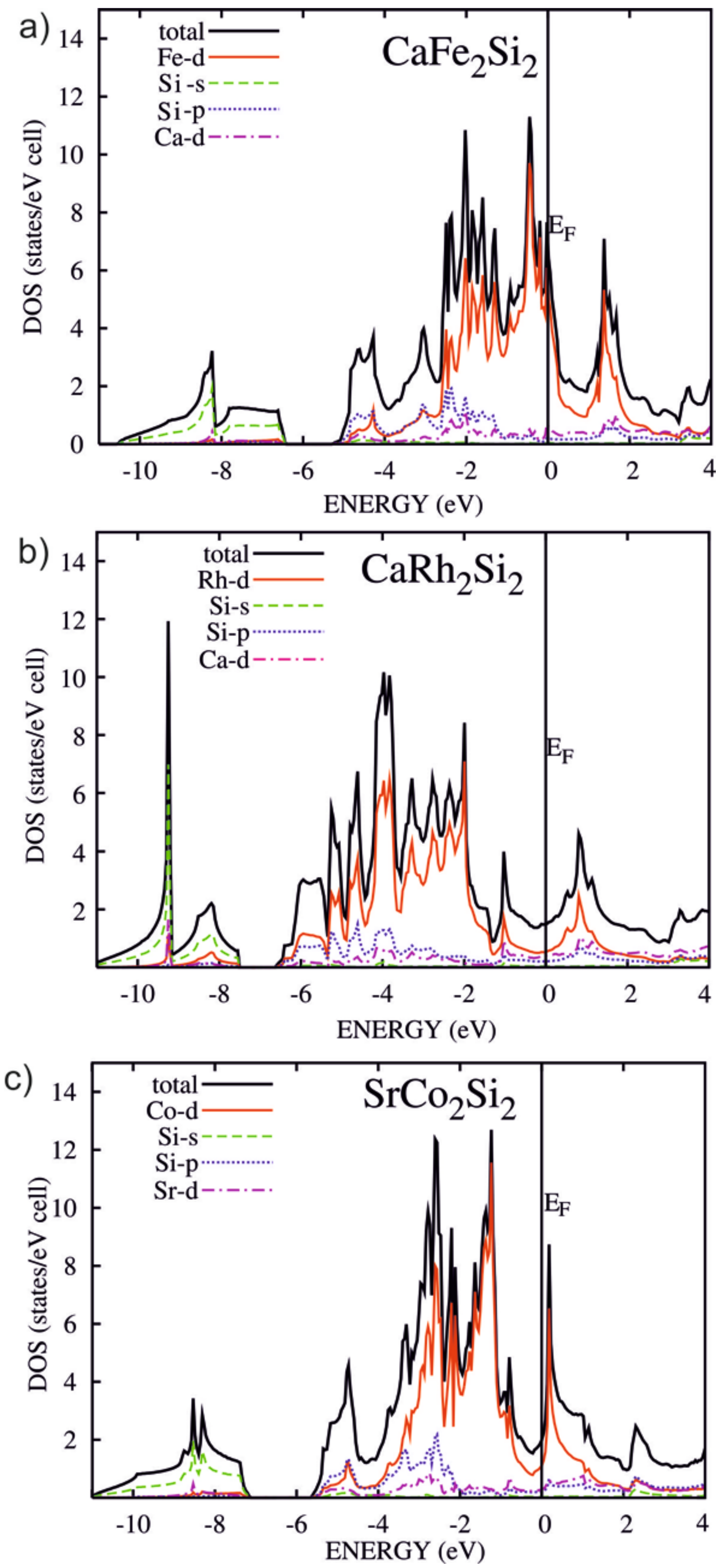

**Figure 2.** DOS and projected DOS calculated for a) $CaFe_2Si_2$, a) $CaRh_2Si_2$ and c) $SrCo_2Si_2$. The energy zero is taken at the Fermi level (for comparison the DOS for $CaCo_2Si_2$ is presented in Supporting Information, Figure S8).

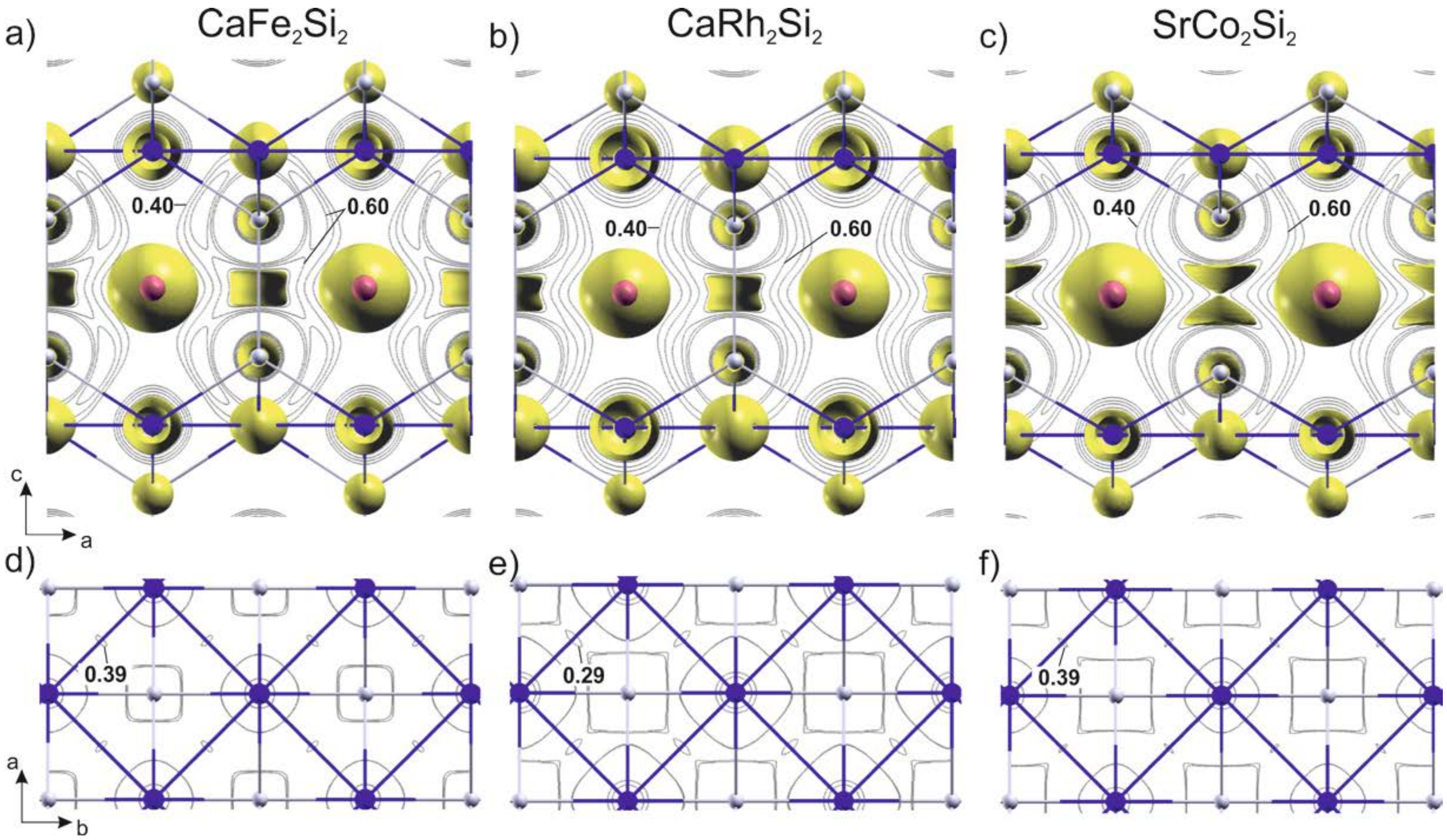

**Figure 3**. Topology of the ELF for $CaFe_2Si_2$, $CaRh_2Si_2$ and $SrCo_2Si_2$ calculated from the all-electron density (TB-LMTO-ASA). A 3D ELF plot with isosurface at $\eta = 0.72$ and contour line diagrams of the ELF in the *ac* plane present the Si-Si bonding in $CaFe_2Si_2$ (a), $CaRh_2Si_2$ (b), and lone electron pairs at the Si atoms in $SrCo_2Si_2$ (c). Contour line diagrams of the ELF in the *ab* plane of $CaFe_2Si_2$ (d), $CaRh_2Si_2$ (e), and $SrCo_2Si_2$ (f) emphasise weak *T*-*T* interactions. For the clarity only the regions of $\eta$ = 0.39 - 040, $\eta$ = 0.285 - 0.29 and $\eta$ = 0.39 - 040 for $CaFe_2Si_2$, $CaRh_2Si_2$ and $SrCo_2Si_2$, respectively, were chosen.

## Supporting Information

# Synthesis and structure of $SrCo_2Si_2$ and $CaRh_2Si_2$ - isoelectronic variants of the parent superconductors $\mathit{Ae}Fe_2As_2$ and study of the influence of the valence electron count in $CaFe_{2-x}Rh_xSi_2$

**Viktor Hlukhyy, Andrea V. Hoffmann, Thomas F. Fässler**

**Table S1** Lattice parameters, atomic distances and tetrahedral angles ($\alpha$) of $AeT_2Ge_2$ ($Ae$ = Ca-Ba, $T$ = Mn-Zn) compounds ($ThCr_2Si_2$-type structure, $I4/mmm$, Z = 2).

| **Compound** | ***a* / Å** | ***c* / Å** | ***V* / Å$^3$** | ***c*/*a*** | ***d*(*T*-*T*) / Å** | ***d*(*T*-Ge) / Å** | ***d*(Ge-Ge) / Å** | **$\alpha$ / °** | **Ref.** |
|---|---|---|---|---|---|---|---|---|---|
| $CaMn_2Ge_2$ | 4.17 | 10.88 | 189.2 | 2.61 | 2.95 | 2.53 | 2.59 | 111.3 | [17] |
| $CaCo_2Ge_2$ | 4.00 | 10.33 | 165.3 | 2.58 | 2.83 | 2.36 | 2.67 | 116.0 | [17] |
| $CaNi_2Ge_2$ | 3.99 | 9.67 | 153.8 | 2.43 | 2.82 | 2.31 | 2.49 | 119.1 | [41] |
| $CaCu_2Ge_2$ | 4.14 | 10.23 | 175.3 | 2.47 | 2.93 | 2.45 | 2.48 | 114.9 | [23] |
| $CaZn_2Ge_2$ | 4.21 | 10.85 | 192.3 | 2.58 | 2.98 | 2.57 | 2.47 | 109.9 | [23] |
| $SrMn_2Ge_2$ | 4.30 | 10.91 | 201.7 | 2.54 | 3.04 | 2.58 | 2.62 | 113.2 | [17] |
| $SrCo_2Ge_2$ | 4.08 | 10.65 | 177.3 | 2.61 | 2.89 | 2.37 | 2.90 | 118.5 | [17] |
| $SrNi_2Ge_2$ | 4.17 | 10.25 | 178.2 | 2.46 | 2.95 | 2.38 | 2.83 | 122.3 | [17] |
| $SrCu_2Ge_2$ | 4.28 | 10.31 | 188.9 | 2.41 | 3.03 | 2.47 | 2.70 | 120.3 | [17] |
| $SrZn_2Ge_2$ | 4.37 | 10.61 | 202.6 | 2.43 | 3.09 | 2.60 | 2.50 | 114.7 | [17] |
| $BaMn_2Ge_2$ | 4.47 | 10.99 | 219.6 | 2.46 | 3.16 | 2.62 | 2.75 | 116.8 | [17] |
| $BaCo_2Ge_2$ | 4.09 | 11.78 | 196.6 | 2.88 | 2.89 | 2.35 | 3.55 | 120.4 | [3] |
| $BaNi_2Ge_2$ | 4.27 | 11.25 | 204.9 | 2.64 | 3.02 | 2.35 | 3.64 | 130.0 | [53] |

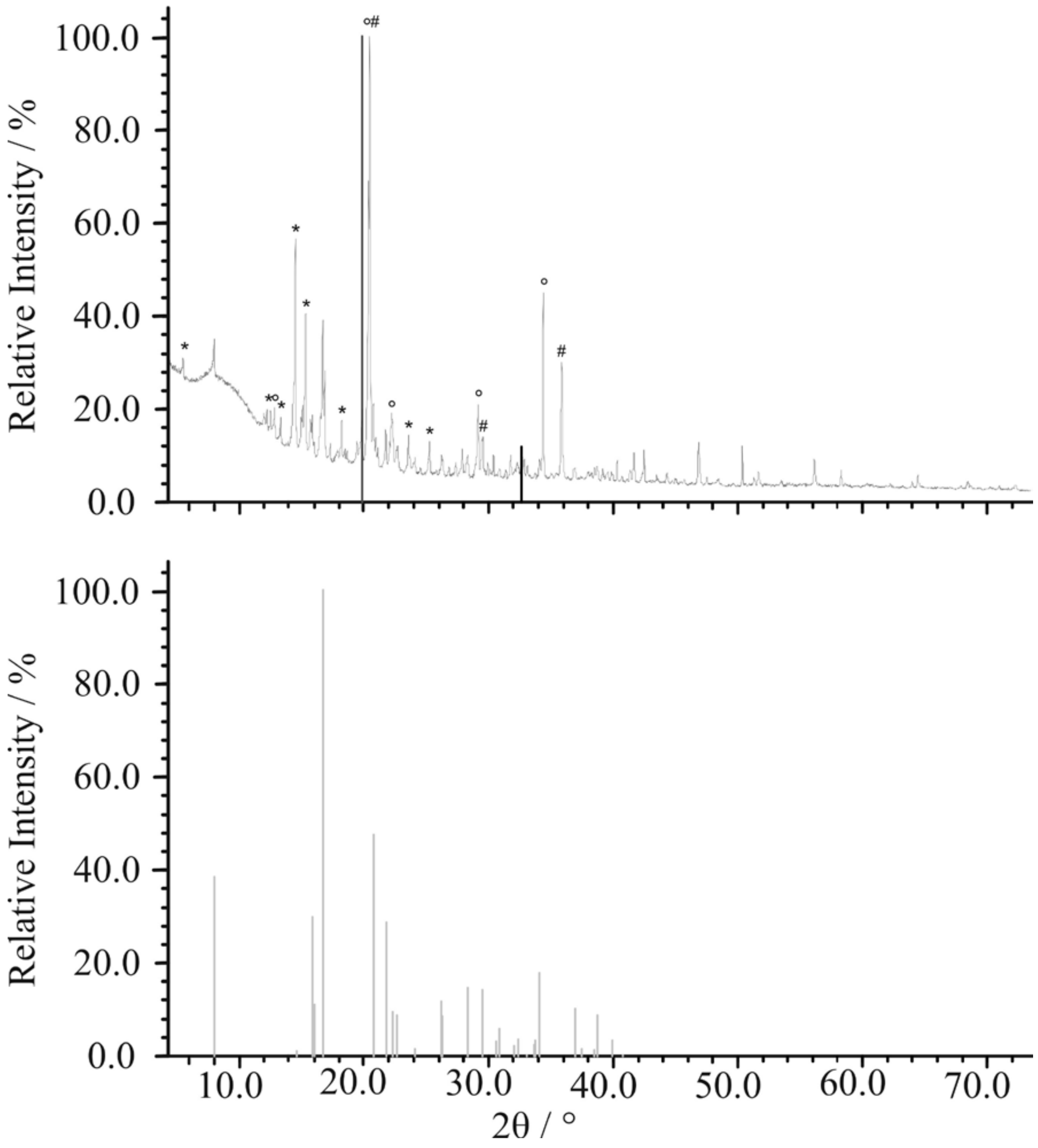

**Figure S1.** X-ray powder diffraction pattern of $CaFe_2Si_2$ sample. Theoretical calculated pattern of $CaFe_2Si_2$ with $ThCr_2Si_2$-type structure (light gray lines) indicated in the diffractogram. The reflexes marked with the dark line at 19.8 ° and 33.65 ° belong to cubic diamond which was used as an internal standard. Reflexes marked with * belong to $Ca_5Si_3$ [54], reflexes marked with ° belong to FeSi[55] and reflexes marked with # belong to $Fe_3Si$ [56]. (Mo-MythenK-PSD)

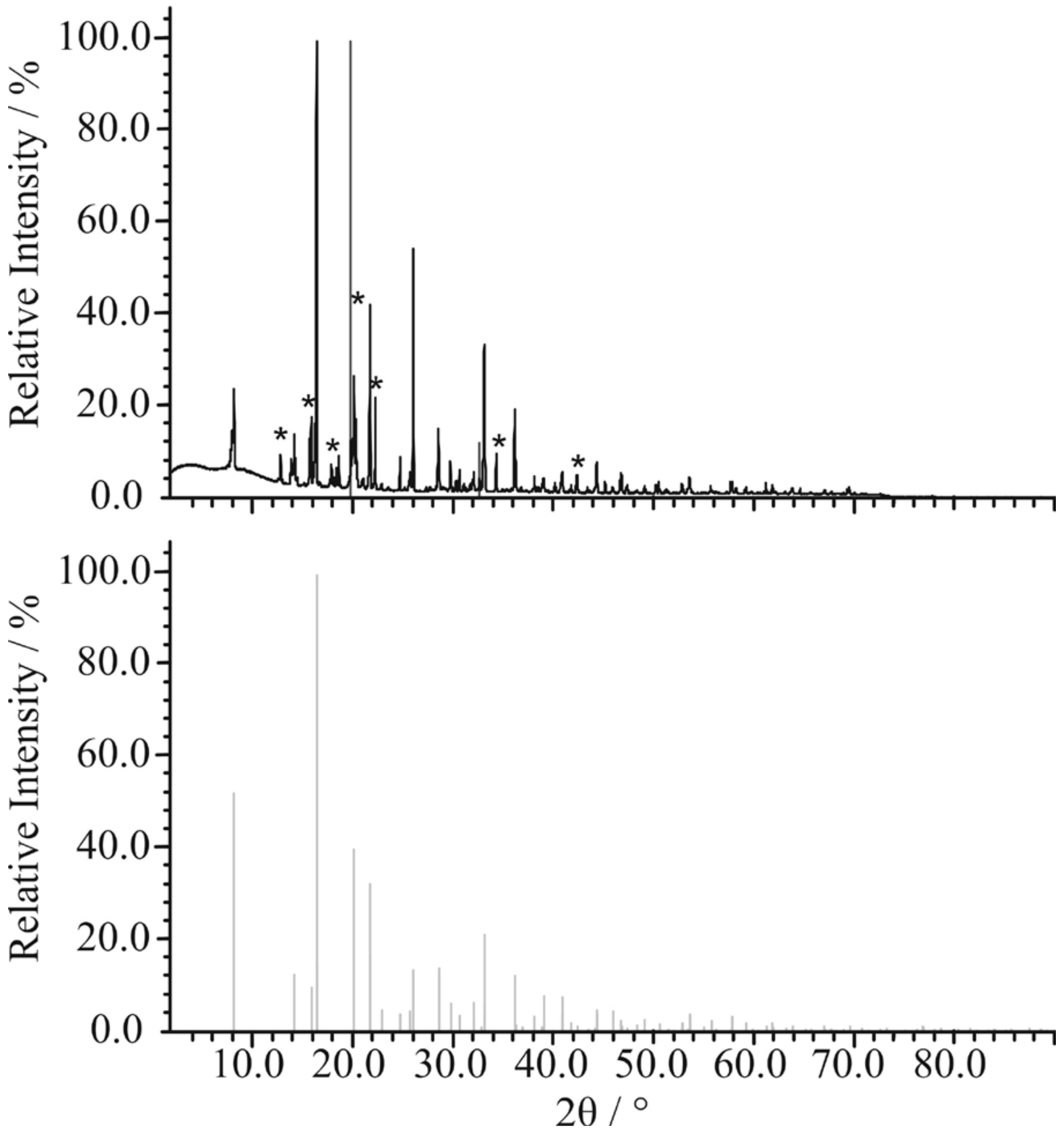

**Figure S2.** X-ray powder diffraction pattern of $CaFe_{0.66(3)}Rh_{1.34(3)}Si_2$ sample. Theoretically calculated pattern of $CaFe_{0.66(3)}Rh_{1.34(3)}Si_2$ with $ThCr_2Si_2$-type structure (light gray lines) indicated in the diffractogram. The reflex marked with the dark line at 19.8 ° and 33.65 ° belong to cubic diamond which was used as an internal standard. Reflexes marked with * belong to FeSi [55]. (Mo-MythenK-PSD)

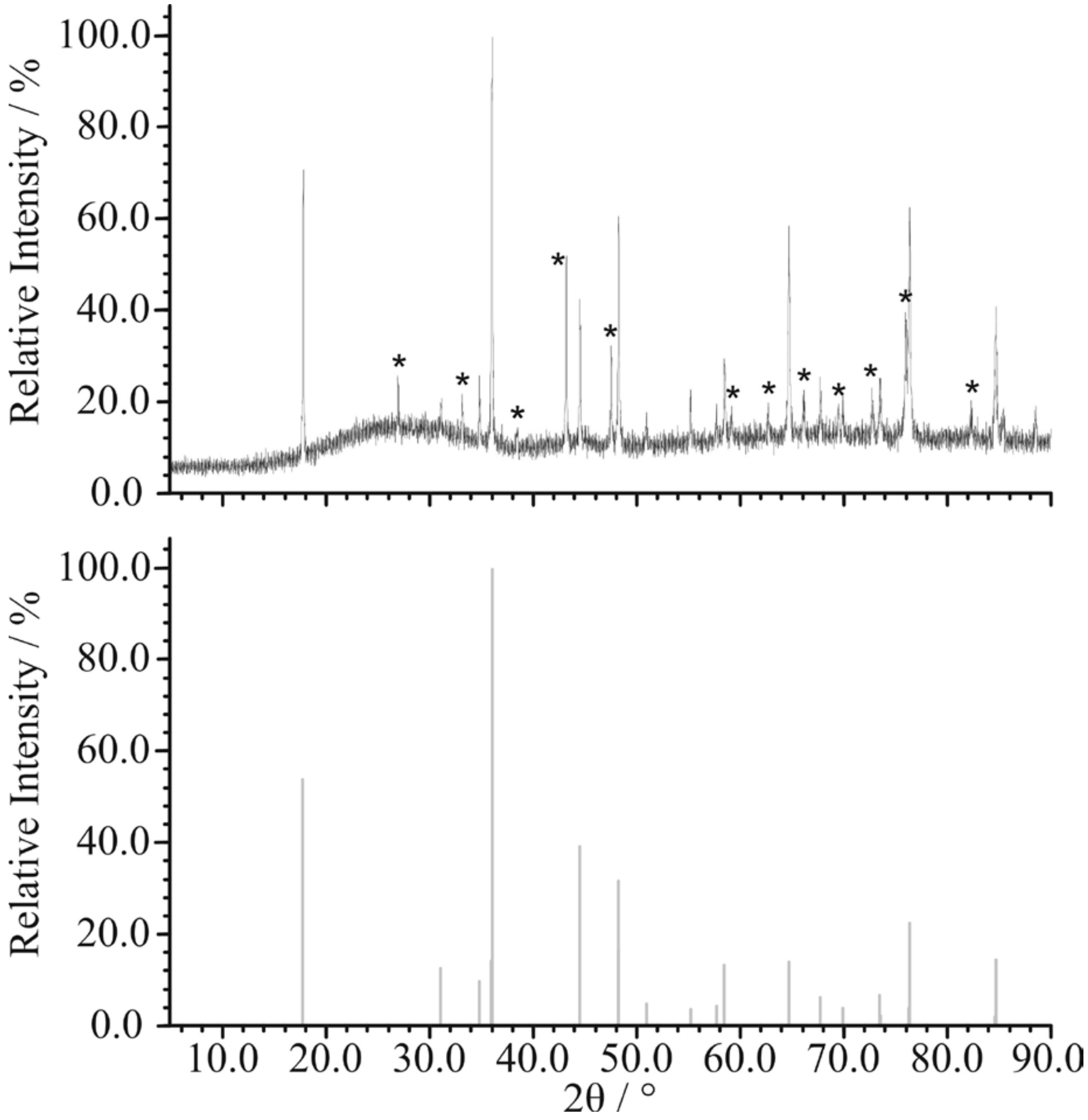

**Figure S3.** X-ray powder diffraction pattern of $CaRh_2Si_2$ sample. Theoretical calculated pattern of $CaRh_2Si_2$ with $ThCr_2Si_2$-type structure (light gray lines) indicated in the diffractogram. Reflexes marked with * belong to RhSi [57]. (L-PSD)

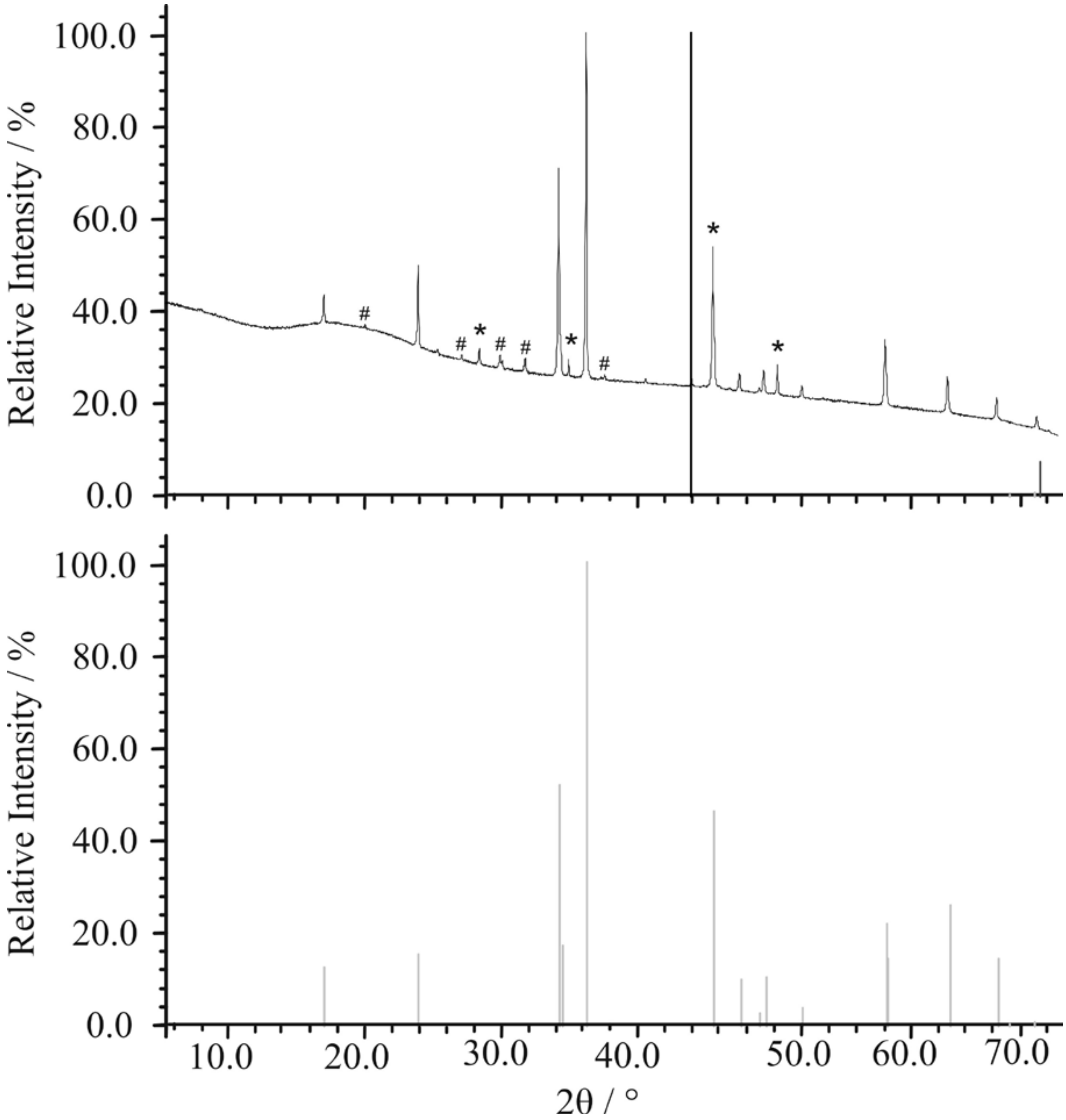

**Figure S4.** X-ray powder diffraction pattern of $SrCo_2Si_2$ sample made with the resistance furnace. Theoretically calculated pattern of $SrCo_2Si_2$ with $ThCr_2Si_2$-type structure (light gray lines) indicated in the diffractogram. Reflexes marked with * belong to CoSi [58] and reflexes marked with # belong to SrSi [59]. (L-PSD)

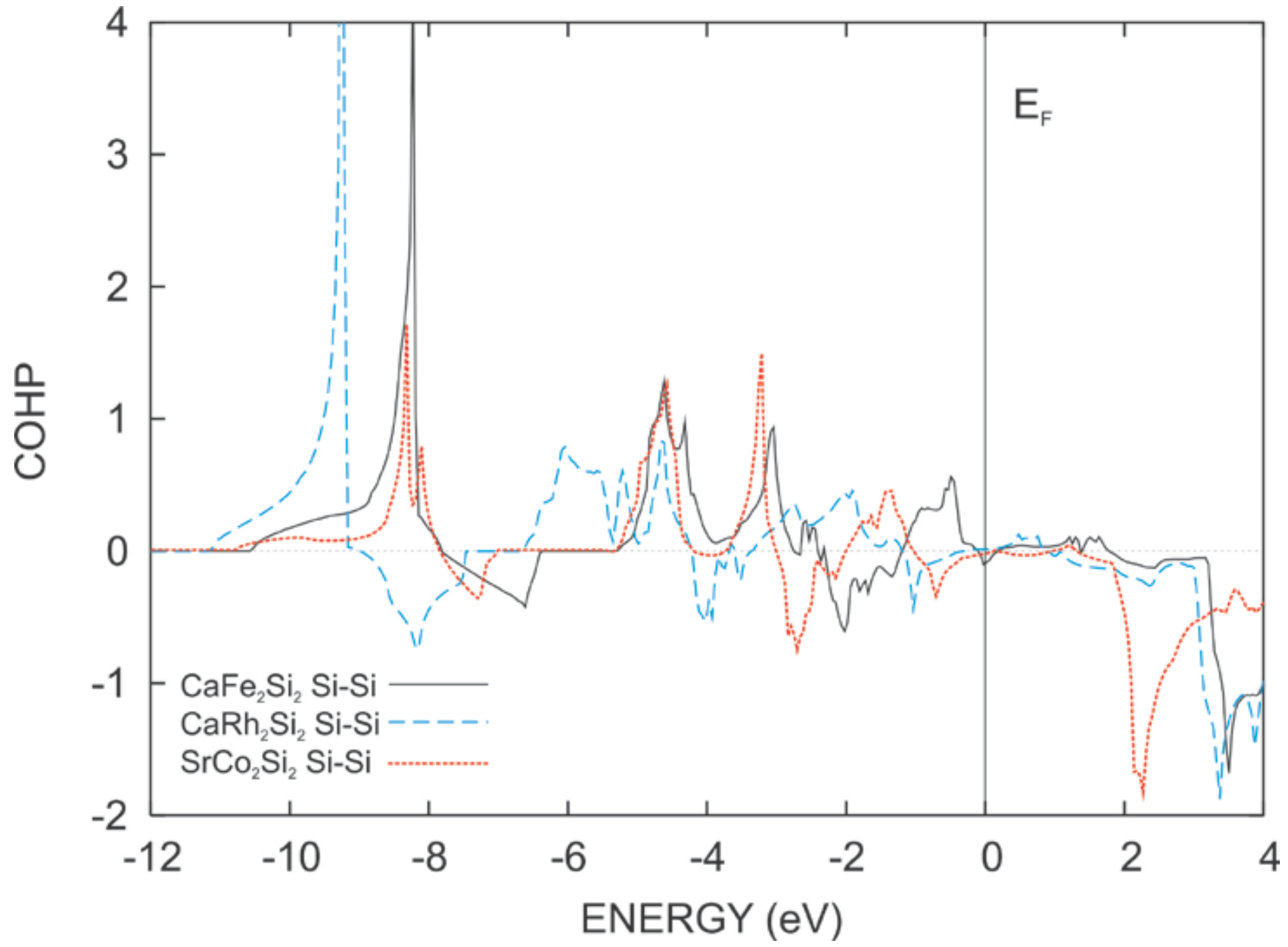

**Figure S5.** Crystal orbital Hamiltonian populations (COHP) curves for the Si-Si bonds in $CaFe_2Si_2$, $CaRh_2Si_2$ and $SrCo_2Si_2$.

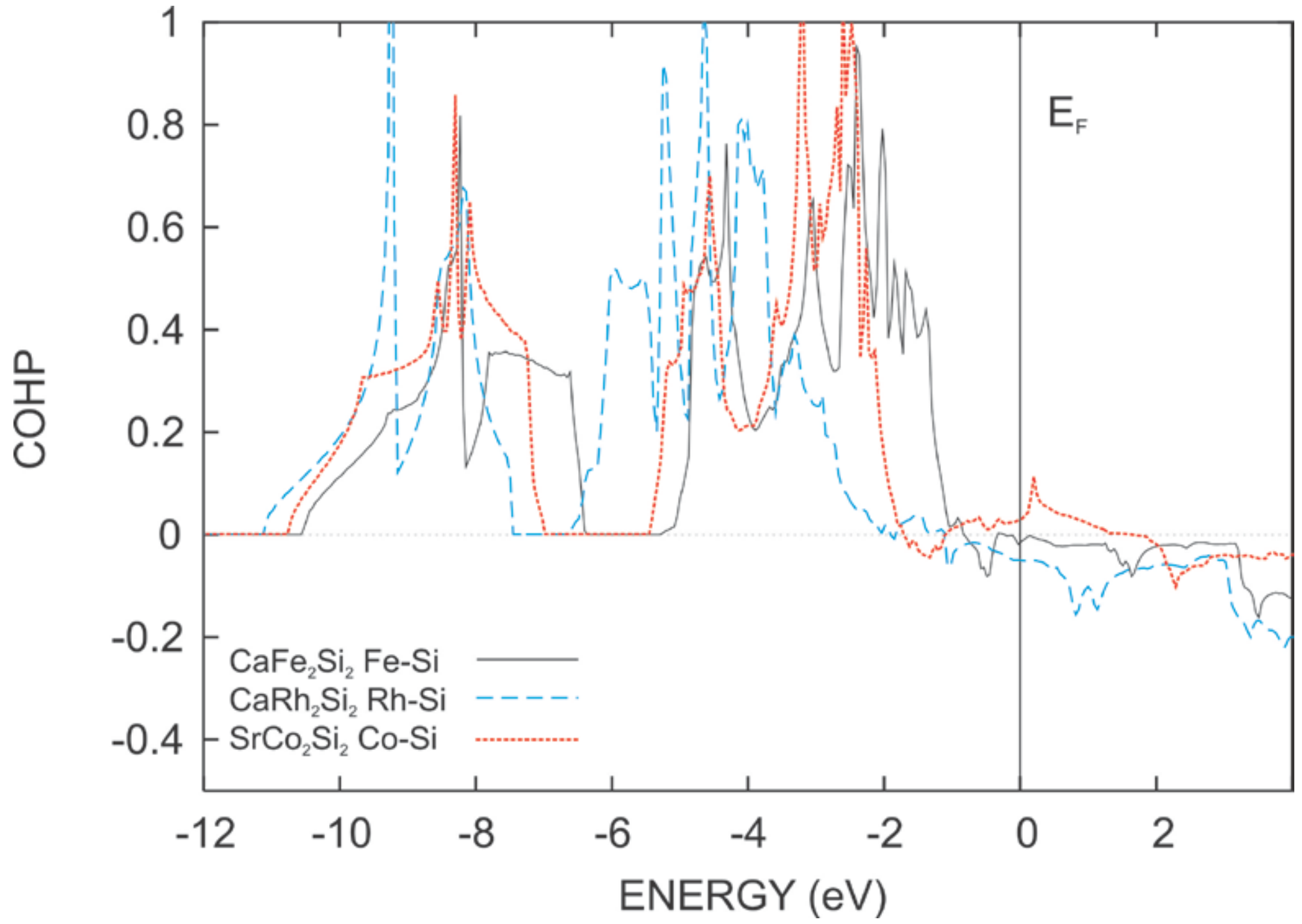

**Figure S6.** Crystal orbital Hamiltonian populations (COHP) curves for the *T*-Si bonds in $CaFe_2Si_2$, $CaRh_2Si_2$ and $SrCo_2Si_2$.

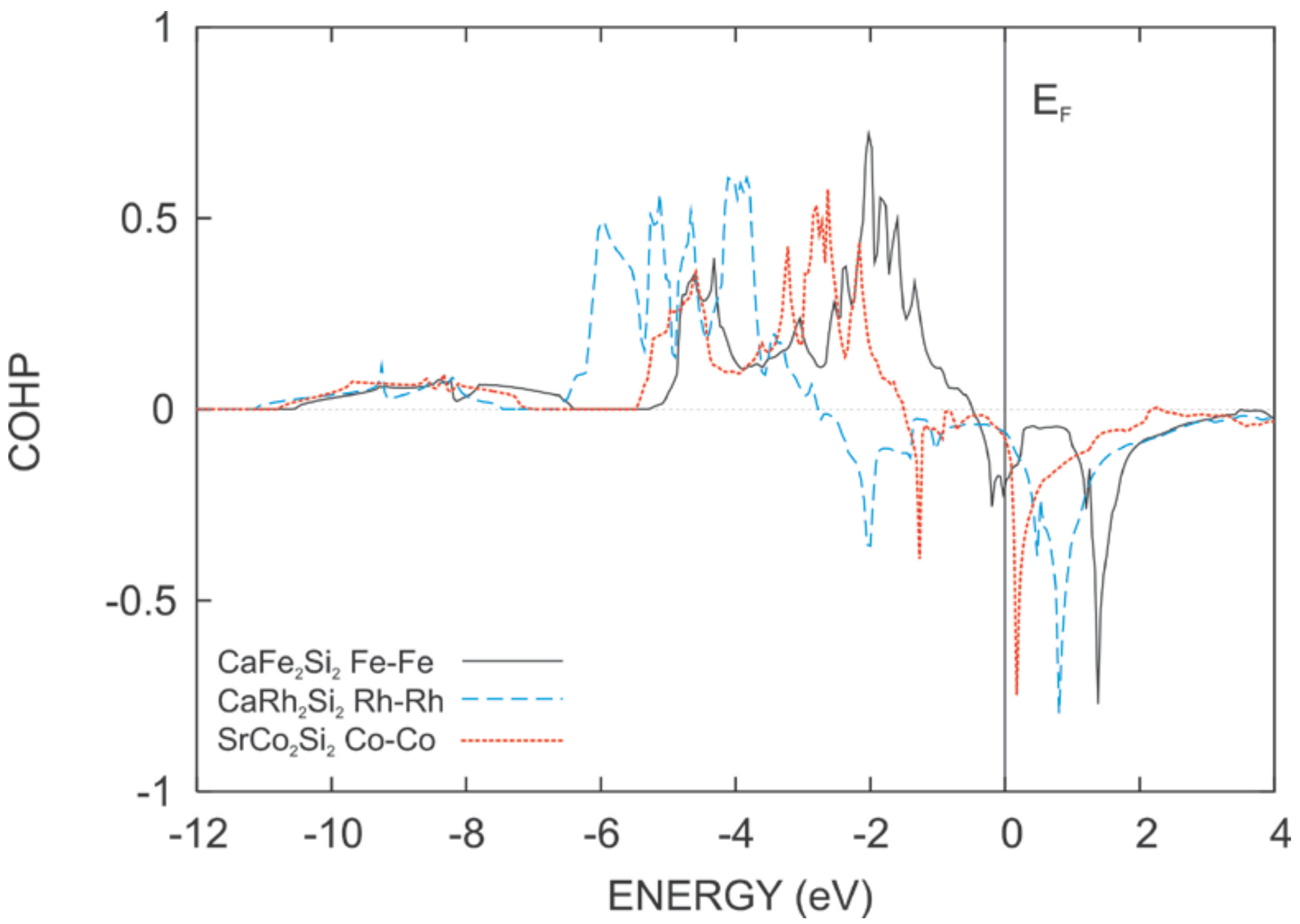

**Figure S7.** Crystal orbital Hamiltonian populations (COHP) curves for the *T*-*T* bonds in $CaFe_2Si_2$, $CaRh_2Si_2$ and $SrCo_2Si_2$.

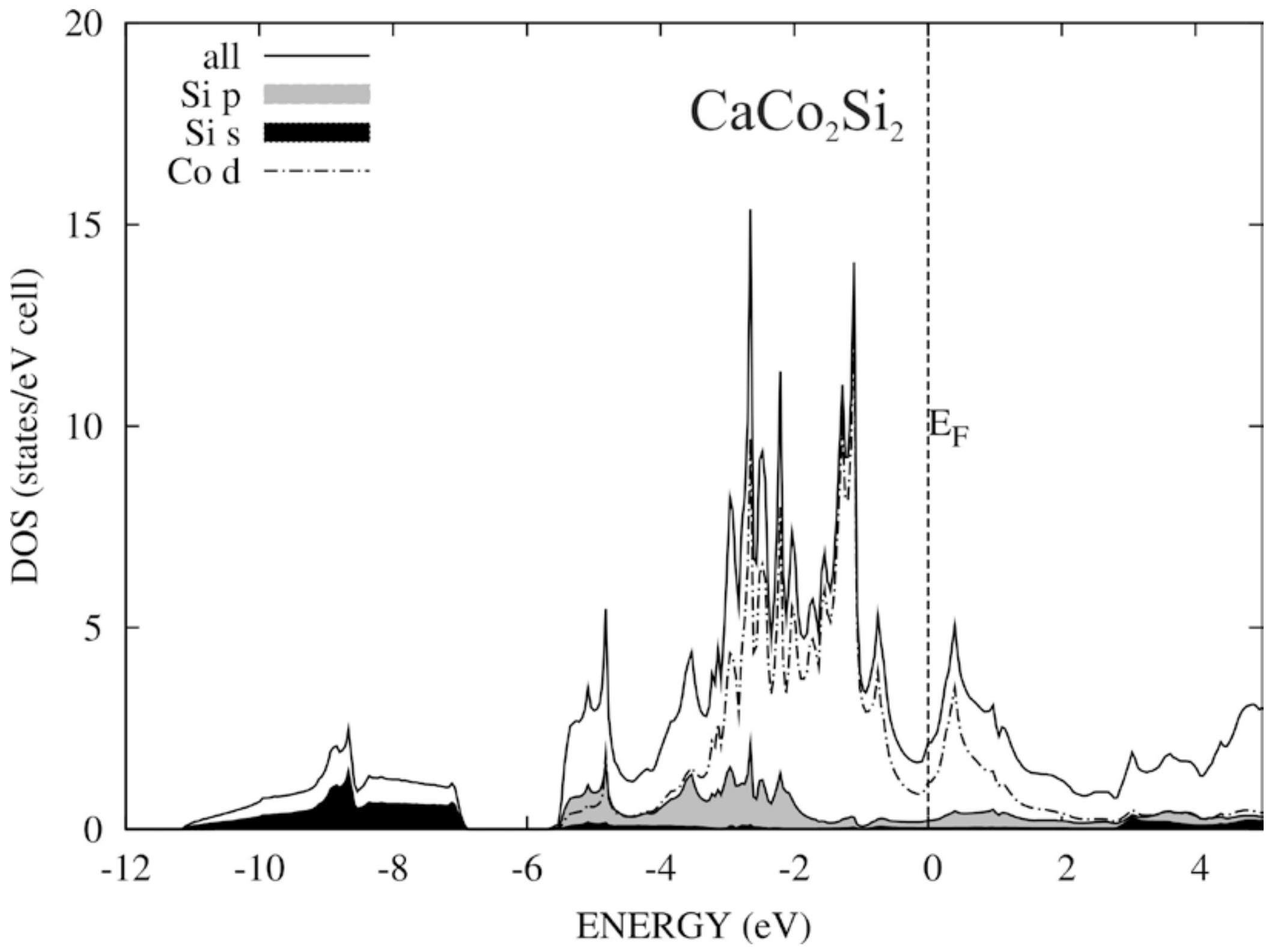

**Figure S8.** Total DOS and partial DOS calculated for $CaCo_2Si_2$. Taken from [3].

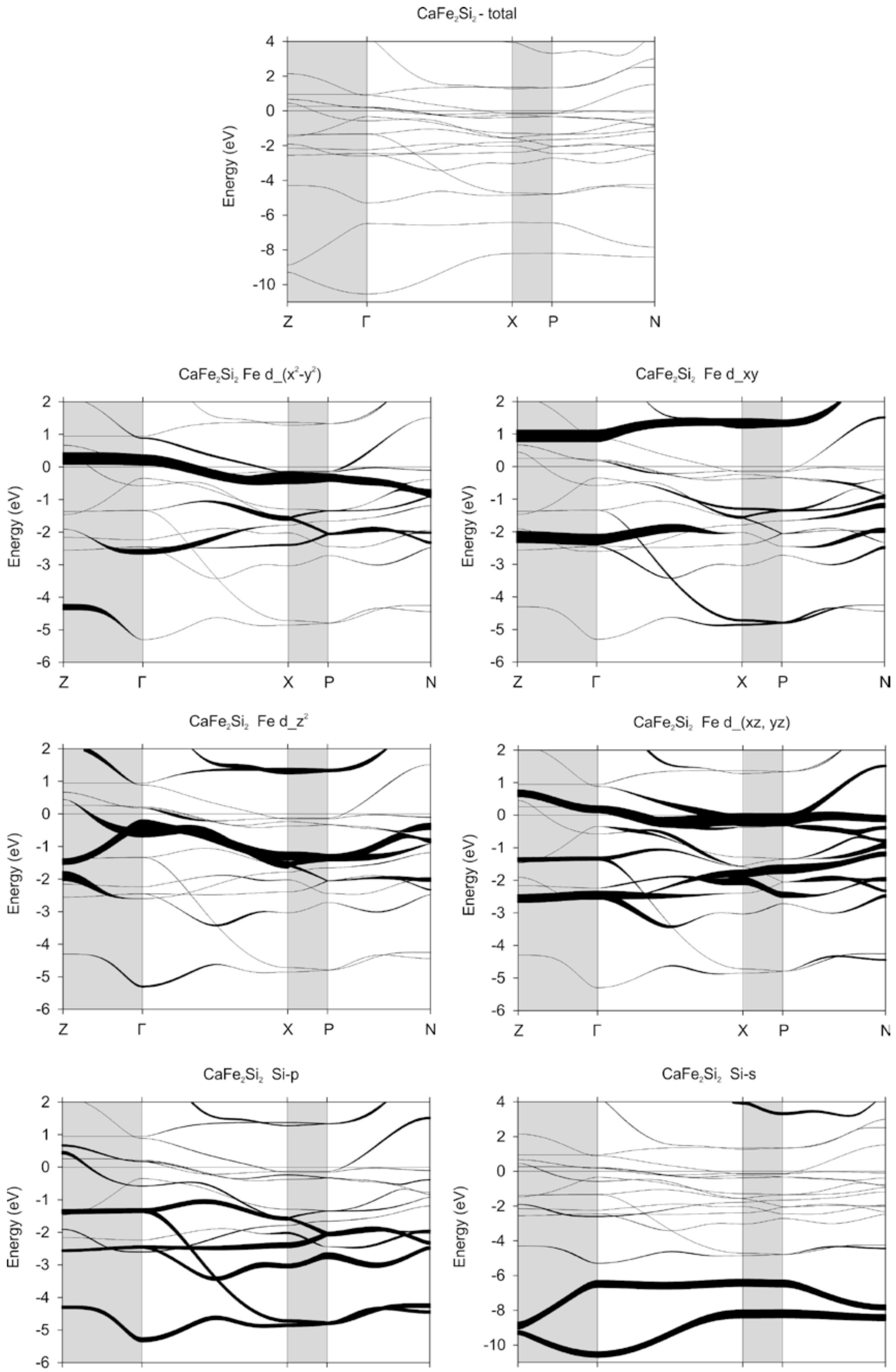

**Figure S9.** Band structure for $CaFe_2Si_2$ in the range from −6 eV to 2 eV including fatbands for Fe- $d_{x^2-y^2}$, $d_{xy}$, $d_{z^2}$, $d_{xz,yz}$, Si- p, s orbitals.

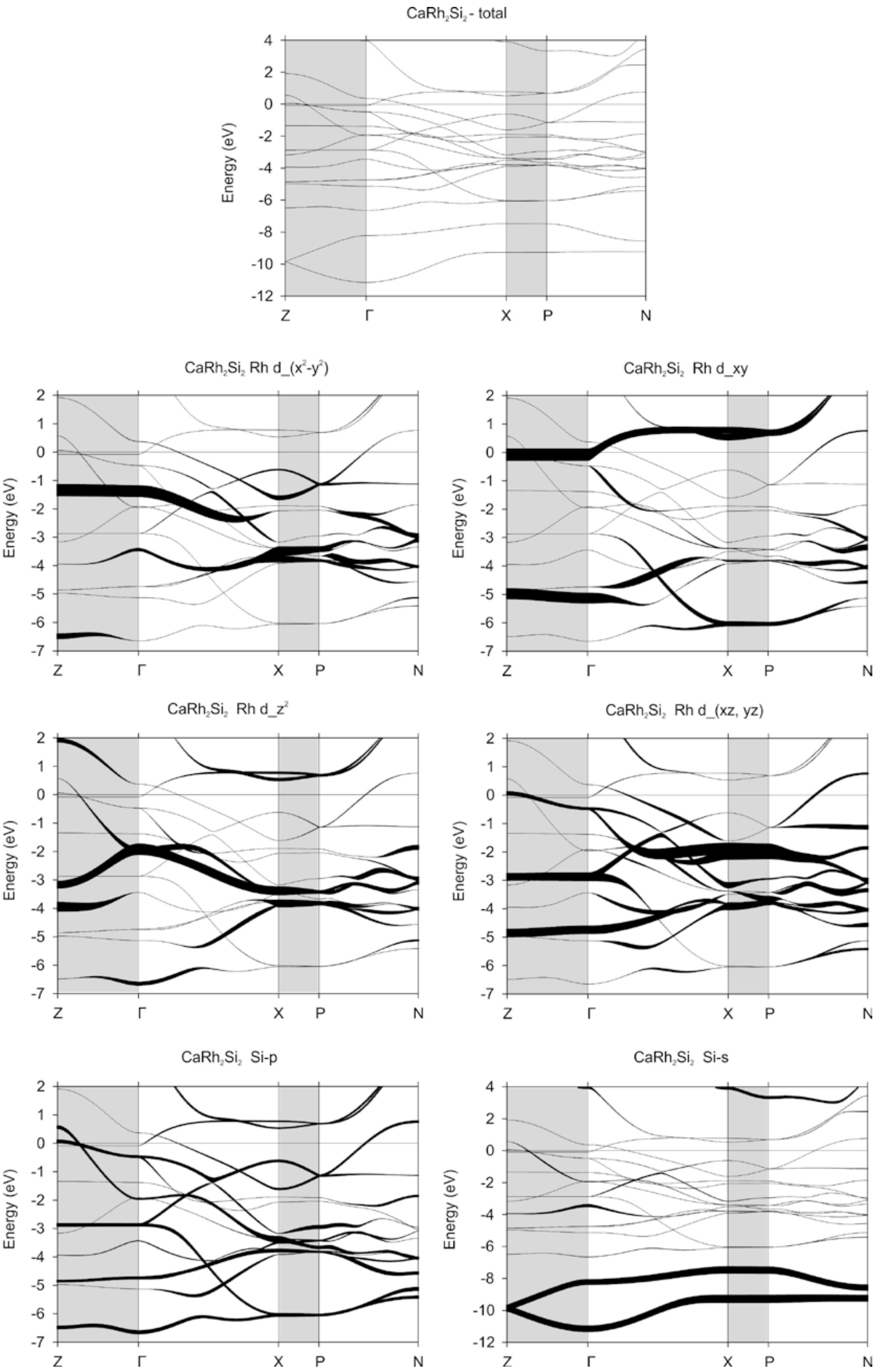

**Figure S10.** Band structure for $CaRh_2Si_2$ in the range from −6 eV to 2 eV including fatbands for Rh- $d_{x^2-y^2}$, $d_{xy}$, $d_{z^2}$, $d_{xz,yz}$, Si- p, s orbitals.

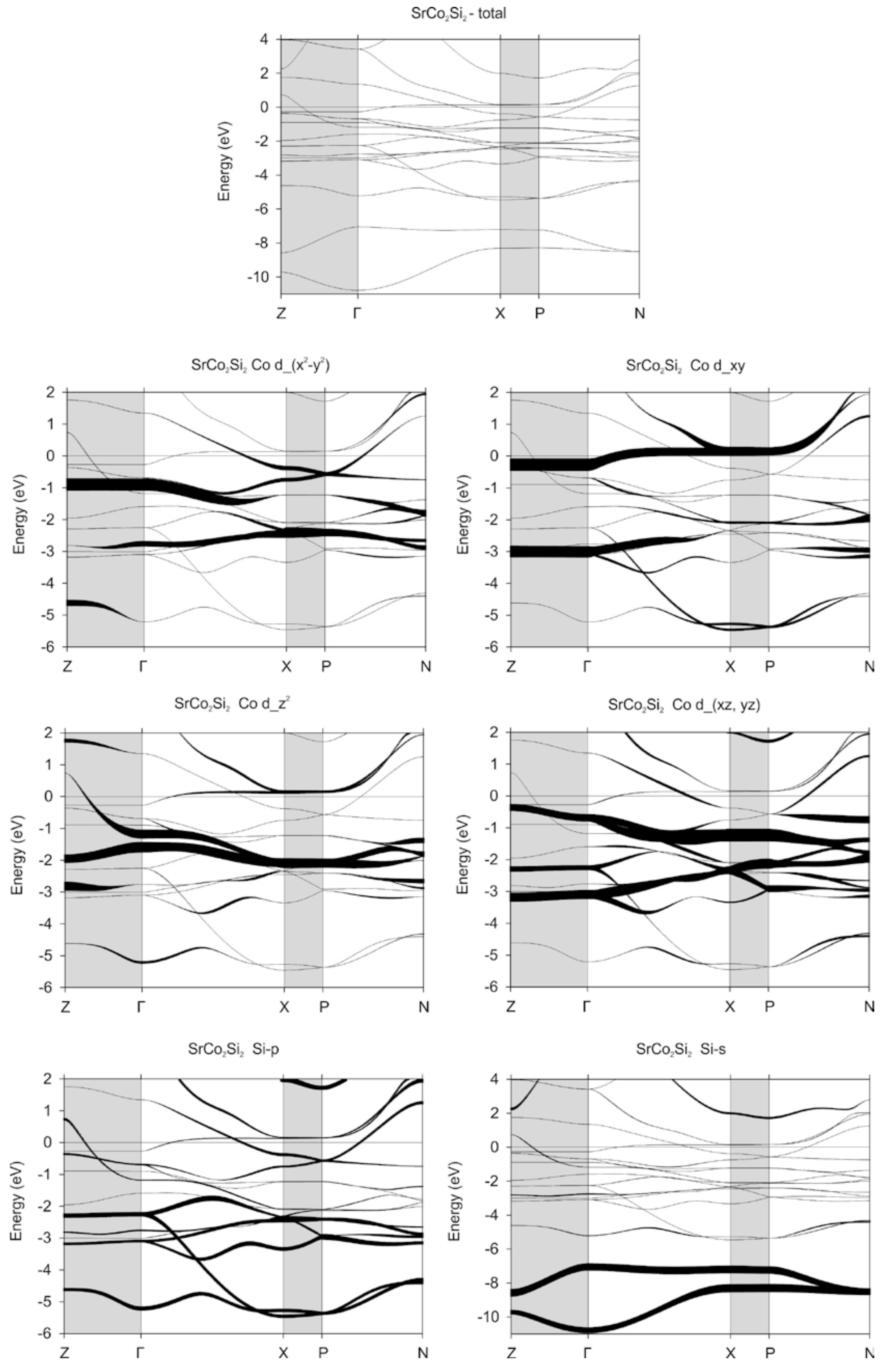

**Figure S11.** Band structure for $SrCo_2Si_2$ in the range from −6 eV to 2 eV including fatbands for Co- $d_{x^2-y^2}$, $d_{xy}$, $d_{z^2}$, $d_{xz,yz}$, Si- p, s orbitals.